\documentclass[12pt,preprint]{aastex}

\shorttitle{Dwarf Galaxy Starburst Fractions}
\shortauthors{Lee et al.}

\begin{document}

\title{Dwarf Galaxy Starburst Statistics in the Local Volume}


\author{Janice C. Lee\altaffilmark{1,2,3}, Robert C. Kennicutt, Jr.\altaffilmark{2,4},
Jos\'e G. Funes, S.J.\altaffilmark{5,6},\\
Shoko Sakai\altaffilmark{7}, Sanae Akiyama\altaffilmark{2}}

\altaffiltext{1}{Present Address: Carnegie Observatories, 813 Santa Barbara
Street, Pasadena, CA  91101; jlee@ociw.edu}
\altaffiltext{2}{Steward Observatory, University of Arizona, Tucson, AZ 85721}
\altaffiltext{3}{Hubble Fellow}
\altaffiltext{4}{Institute of Astronomy, University of Cambridge,
Madingley Road, Cambridge CB3 0HA, UK}
\altaffiltext{5}{Vatican Observatory, Steward Observatory, University of Arizona, Tucson, AZ 85721}
\altaffiltext{6}{Visiting Astronomer, Cerro Tololo Inter-American Observatory.
CTIO is operated by AURA, Inc. under contract to the National Science
Foundation.}
\altaffiltext{7}{Division of Astronomy and Astrophysics, University
of California, Los Angeles, Los Angeles, CA, 90095-1547}

\begin{abstract}
An unresolved question in galaxy evolution is whether the star formation histories of low mass systems are preferentially dominated by starbursts or modes that are more quiescent and continuous. Here, we quantify the prevalence of global starbursts in dwarf galaxies at the present epoch, and infer their characteristic durations and amplitudes. The analysis is based on the H$\alpha$ component of the 11 Mpc H$\alpha$ UV Galaxy Survey (11HUGS), which is providing H$\alpha$ and GALEX UV imaging for an approximately volume-limited sample of $\sim$300 star-forming galaxies within 11 Mpc. We first examine the completeness properties of the sample, and then directly tally the number of bursting dwarfs and compute the fraction of star formation that is concentrated in such systems.  To identify starbursting dwarfs, we use an integrated
H$\alpha$ EW threshold of 100\AA, which corresponds to a stellar birthrate of $\sim$2.5, and also explore the use of empirical starburst
definitions based on $\sigma$-thresholds of the observed
logarithmic EW distributions.  Our results are robust to the exact choice of the threshold, and are consistent with a picture where dwarfs that are currently experiencing massive global bursts are just the $\sim$6\% tip of a low-mass galaxy iceberg.  Moreover, bursts are only responsible for about a quarter of the total star formation in the overall dwarf population, so the majority of stars in low-mass systems are not formed in this mode today. Spirals and irregulars devoid of H$\alpha$ emission are rare, indicating that the complete cessation of star formation generally does not occur in such galaxies and is not characteristic of the inter-burst state, at least for the more luminous systems with $M_B<-15$.  The starburst statistics presented here directly constrain the duty cycle and the average burst amplitude under the simplest assumptions where all dwarf irregulars share a common star formation history and undergo similar bursts cycles with equal probability.  Uncertainties in such assumptions are discussed in the context of previous work.

\end{abstract}





\section{Introduction}
Do cycles of violent, intense, but short-lived global bursts constitute the dominant mode of star formation in low-mass galaxies?  This question was originally raised over thirty years ago, in connection with the discoveries of dwarf galaxies that resemble ``isolated extragalactic HII regions'' (Sargent \& Searle 1970).  These systems, now commonly referred to as the blue compact dwarfs (BCDs) or HII galaxies, were first identified in Zwicky's (1964, 1966) catalogs of ``compact'' and ``eruptive'' galaxies, and in Markarian's (1967, 1969a,b) survey for objects with large ultraviolet excesses.   Star formation histories punctuated by strong ``flashes'' were then eventually proposed as the most likely solution to the puzzle presented by the anomalously blue colors and low gas-phase metal abundances observed in the lowest luminosity members of these samples (Searle \& Sargent 1972; Searle, Sargent \& Bagnuolo 1973; Huchra 1977b).  

Since then, the study of starbursts in dwarf galaxies has significantly grown due to the gradual recognition that these events may have a profound impact on systems with shallow potential wells.  
For example, starburst episodes have been invoked, although with much ensuing debate, as an agent which transforms gas-rich dwarf irregulars
into gas poor dwarf ellipticals through gas consumption and expulsion (e.g., Vader 1986; Dekel \& Silk 1986; Skillman \& Bender 1995; Papaderos et al. 1996; Marlowe et al. 1999; van Zee et al. 2001; Gil de Paz \& Madore 2005).  The stellar winds and supernovae produced by starbursts are argued to drive metal-enriched winds which may escape the haloes of low-mass galaxies and pollute the intergalactic medium (IGM) (e.g., Marlowe et al. 1995; Mac Low \& Ferrara 1999; Martin 1999; Garnett 2002).  Starbursts, therefore, are also implicated in the possibly related phenomenon of the observed decrease of the effective yield and metallicity with decreasing luminosity (e.g., Skillman et al. 1989; Richer \& McCall 1995; J. Lee et al. 2004), rotational velocity (Garnett 2002) and stellar mass (Tremonti et al. 2004; H. Lee et al. 2006).  Finally, bursty behavior in dwarf galaxies is a key feature of numerical models which attempt to reasonably incorporate the effects of supernova feedback (e.g., Pelupessy et al. 2004; Stinson et al. 2007). 

Open questions of fundamental importance to all these issues deal with the prevalence of starbursts in dwarfs.  We must ask: {\em What are the characteristic durations, frequencies and amplitudes of the starburst cycles which determine their efficacy as critical sinks of fuel and impulsive sources of disruptive energy?  What is the mass fraction of stars formed during the burst phases?  How do these parameters vary over cosmic time?  Moreover, are all low mass galaxies are equally prone to bursting episodes, or rather does the starburst mode only operate in a particular sub-set of the population?}

In order to fully answer these questions, we would require a statistically complete sample of dwarf galaxies that span the total range of star formation activities, from those systems which are presently undergoing a starburst event, to those which are in a period of relative quiescence.  
Star formation histories (SFHs) would be needed for each of galaxies in this sample, and the temporal sampling of the SFHs must be fine enough to resolve a burst cycle.  The typical modes of global star formation can then be characterized.  If violent fluctuations in the star formation rate (SFR) are evident for only a limited subset of galaxies, then one can search for commonalities in the physical properties among those objects that may distinguish them from non-bursting systems.

The trouble is that we typically cannot follow the SFHs of individual systems back through cosmic time.  The exception is for the galaxies in and around the Local Group.  The stellar populations of these systems can be resolved and observed to sufficient depth such that color-magnitude diagrams (CMDs) can be used to reconstruct detailed SFHs.  The aggregate of such studies for dwarf irregulars have shown that some have had only modest fluctuations in their star formation rates (e.g., Tosi et al. 1991; Greggio et al. 1993; Marconi et al. 1995; Aparicio et al. 1997a, b; Dohm-Palmer et al 1998; Gallagher et al. 1998; Dolphin et al. 2005).  For example, Dohm-Palmer et al. (1998) derived the recent SFHs of four Local Group dwarf irregulars (Sextans A, Pegasus DIG, Leo A and GR 8) using main sequence and blue helium burning (supergiant) stars, and found that there were no bursts (no star formation episodes that deviated from the average activity by more than a factor of $\sim$3) in these systems over the past 500 Myrs.  Assuming a burst duration of 100 Myr, they then inferred that the total time spent in the burst state must be less than $\sim$5\%, since there was the potential of observing 20 burst episodes in their data, but zero were observed.  Although this is a potentially powerful technique for constraining the duty cycle and the prevalence of starbursts in the evolution of dwarfs, its application is currently limited by small number statistics.  Present capabilities only allow the stellar populations of galaxies that are within a few Mpc to be resolved to requiste depth (e.g., Dalcanton et al. 2008, Weisz et al. 2008).  The next generation of extremely large 30-m class telescopes should enable the application of this technique to a statistical sample of dwarf galaxies out to $\sim$15 Mpc, but in the meantime, we require an alternate strategy.

Another approach that has long been the staple of galaxy evolution studies uses the present-day {\it integrated} properties of galaxies, in particular the $UBV$ colors and H$\alpha$-based SFRs, to infer the SFHs (e.g., Tinsley 1968, 1972; Searle, Sargent \& Bagnuolo 1973; Gallagher, Hunter \& Tutukov 1984; Kennicutt 1998 and references therein; van Zee 2001; Bruzual \& Charlot 2003 and references therein).  Although this method is coarser in its temporal resolving power, it allows for the analysis of the larger sample that is needed, and so, is complementary to the CMD studies.  The picture that has emerged from such work on dwarf irregulars is that the optical colors and current SFRs of the majority of the population can be explained by SFHs characterized by constant, but relatively low SFRs (e.g. Hunter \& Gallagher 1985).  Taken together with the CMD results described above, this has led some to argue against a significant starburst phase in the majority of dwarf irregulars (e.g. van Zee 2001).  However, the inherent weakness here is that these studies have generally utilized samples that are only representative, and not statistically complete.  Perhaps the most homogeneous and best characterized samples containing large number of low-luminosity galaxies have been produced by the emission-line and color-selected surveys.  But of course, programs such as the Haro (1956), Markarian (1967, 1969a,b), UCM (Zamorano et al 1994.) and KISS (Salzer et al. 2000) surveys, preferentially select the most strongly star-forming systems by design, and become severely incomplete for galaxies in which the current star formation rate is not elevated relative to the average past activity.  

In this paper, we take a natural next step in constraining the average durations, frequencies and amplitudes of starbursts in low-mass galaxies.  We provide a star formation analysis based on a dwarf galaxy sample whose statistical completeness is well-understood {\it and} which probes a full range of activities, from the starbursting BCDs/HII galaxies to the more difficult to observe quiescent, low surface brightness systems in a given volume.  The starburst duty cycle (the fraction of time that is spent in the burst mode) can be constrained using such a sample since the relative numbers of dwarfs observed in various phases of the cycle will scale with the relative durations of those phases.  Thus, the approach is to directly tally the number of bursting dwarfs and compute the fraction of star formation that is concentrated in these systems.  Although the required calculations themselves are straightforward and essentially reduce to a counting experiment, the challenge lies in the construction of an appropriate sample.  In fact, the difficultly of assembling a robust sample is what has primarily impeded significant progress from being made on this issue since the initial work of Searle \& Sargent (1972).  

Our particular strategy has been to focus on our nearest neighbors and take an approximately volume-limited (and hence dwarf-dominated) star formation inventory of galaxies within 11 Mpc of the Milky Way.  The 11 Mpc H$\alpha$ UV Galaxy Survey (11HUGS) has obtained narrowband H$\alpha$ emission-line imaging and is also collecting GALEX UV imaging for all known spirals and irregular galaxies, as well as star-forming early-type galaxies with $d < 11$ Mpc, $|b|>20\degr$, and $B<$15 mag.  Here, we use data from the completed H$\alpha$ component of the parent survey (Kennicutt et al. 2008), which provides tracers of the SFRs and birthrates (ratio of the present to past average SFR) over the past few million years, to calculate the dwarf galaxy starburst statistics described above. 

We begin by giving an overview of the 11HUGS program and describing the H$\alpha$ flux and EW dataset in \S\ 2.  The completeness properties of the sample with respect to blue absolute magnitude and HI mass are examined.  In \S\ 3, we devise a criterion for quantitatively distinguishing starbursts from more normal systems, and then apply the criterion to calculate the number and star formation fractions of dwarf starburst galaxies.  We note that in this analysis we are mainly interested in identifying global (galaxy-wide) starbursts that occur in low-mass systems, and do not address the highly localized, circumnuclear starbursts observed in more massive galaxies (Kennicutt 1998; Kormendy \& Kennicutt 2004 and references therein). In \S\ 4, our statistics are compared with previous work, and we discuss the implications for the average duty cycle, burst amplitudes and inter-burst state.  We conclude with a summary of our results in \S\ 5.  $H_{\circ}=$75 km s$^{-1}$ Mpc$^{-1}$ is adopted for calculating distance dependent quantities, when more direct distance measurements (e.g., based on standard candles) are not available.

\section{The 11HUGS Dataset}

The goal of the 11 Mpc H$\alpha$ UV Galaxy Survey (11HUGS) is to fill a vital niche in existing multi-wavelength surveys of present-day galaxies with a statistically robust, approximately volume-complete study of our nearest star-forming neighbors.  The dataset consists of snapshots of massive star formation as captured via narrowband H$\alpha$ imaging (recombination emission of gas ionized by O and early-type B stars), as well as GALEX NUV (1500 \AA) and FUV (2300 \AA) imaging (photospheric emission from O \& B stars), which traces star formation over a longer ($\sim$10$^8$ yr) timescale.  Thus, 11HUGS provides a foundation for follow-up resolved studies of the HII region populations, star formation, chemical abundances, and ISM properties of local galaxies.  The scientific scope of 11HUGS is further being expanded with the recent addition of Spitzer IRAC mid-infrared and MIPS far-infrared imaging through the Local Volume Legacy (LVL) collaboration\footnote{http://www.ast.cam.ac.uk/IoA/research/lvls} (Lee et al. 2008; Dale et al. 2008).  The Spitzer data will provide crucial information on the dust properties and old stellar population content of the sample.  Public data releases of our mutli-wavelength imaging have begun through the NASA/IPAC Infrared Science Archive\footnote{http://ssc.spitzer.caltech.edu/legacy/lvlhistory.html} (IRSA), with expected completion of data product deliveries by the end of 2009. 

The analysis in this paper is based on the completed H$\alpha$ imaging component of the survey.  Kennicutt et al. (2008; hereafter Paper I) present the resulting integrated H$\alpha$ flux and equivalent width (EW) parent catalog, along with details on the sample selection, observations, and photometry.  In this section, we first give a summary of the dataset, and then examine the completeness limits of the sample.

Our local volume sample is a compilation of all currently known spiral and irregular galaxies within a distance of 11 Mpc, outside of the plane of the Milky Way ($|b|>20\degr$), and brighter than $\sim$15 B magnitudes (N=261).  These limits define the ranges over which the existing catalogs from which the sample is derived (e.g., the Nearby Galaxies Catalog, Tully 1988a) are relatively complete (e.g., Tully 1998b).  Through a combination of new narrowband H$\alpha$+[NII] and $R$-band imaging for 184 of these galaxies, and data compiled from the literature for another 54, integrated H$\alpha$+[NII] fluxes and EWs are available for 91\% of this sample.  The galaxies that remain without H$\alpha$ data are generally southern objects with distances and brightnesses near the sample limits (see Figure 4 in Paper I, and Figure 1b below).  These galaxies drifted into the sample during revisions (occurring after our primary observational campaigns were completed) which incorporated updated distance estimates and photometry.  This is an inherent difficulty associated with efforts to construct a volume-limited sample.  The membership of the sample will be necessarily fluid until accurate distance and photometric measurements are available for all of the galaxies that are within the volume and around its periphery.  As observing time allowed, H$\alpha$ data were also taken for another 123 galaxies within the 11 Mpc volume which were fainter and/or in the zone of avoidance.  Data from the literature is available for 52 such other objects.  The sum of all these galaxies (N=436) comprise the overall dataset published in Paper I.  The work presented below, as well as the follow-on 11HUGS and Spitzer LVL surveys, generally focus on the subsets of the Paper I sample that avoid the Galactic Plane ($|b|>20\degr$).

\subsection{Completeness}\label{completeness}

Ultimately, the 11HUGS sample is a composite of numerous catalogs with diverse selection criteria.  For details on the construction of the sample, the reader is referred to Lee (2006) and Kennicutt et al. (2008).  Here, we estimate the completeness limits of our sample by performing a statistical test described in Rauzy (2001), which is similar to the well-known $V/V_{max}$ test (Schmidt 1968), but does not require the galaxy distribution to be spatially homogeneous.  We also qualitatively compare our number density distributions with independently established B-band luminosity and HI mass functions.  

\subsubsection{The Rauzy $T_C$ Completeness Statistic}

\noindent The Rauzy (2001) $T_C$ completeness statistic is analogous to the $V/V_{max}$ test (Schmidt 1968), but does not rely on assumptions about the spatial homogeneity of the sample.  The method is based on the estimation of the uniform variate $\zeta$, which is a ratio of number densities.  Specifically, if $F(M)$ is the normalized integral of the luminosity function from $-\infty$ to $M$, and $Z$ is the distance modulus $m-M$, then

\begin{equation}
\zeta\equiv\frac{F(M)}{F[M_{lim}(Z)]}
\end{equation}

\noindent where $M_{lim}(Z)$ is the faintest absolute magnitude that a galaxy at a given $Z$ can have and still be visible, given a limiting magnitude $m_{lim}$.

\noindent An estimate of $\zeta$ is given by 

\begin{equation}
\zeta_i\sim\frac{r_i}{n_i+1},
\end{equation}

\noindent where $r_i$ is the number of galaxies with $M\le M_i$ and $Z\le Z_i$, and $n_i$ is the number of galaxies with  $M\le M_{lim}^i$ and $Z\le Z_i$.  If $M$ and $Z$ are uncorrelated, as should be the case for a complete local sample of galaxies, the expectation value of $\zeta$ is 0.5.  The variance is given by

\begin{equation}
V_i=\frac{1}{12}\frac{n_i-1}{n_i+1}.
\end{equation}

\noindent The $T_C$ completeness statistic is then computed as

\begin{equation}
T_C=\frac{\sum\limits_{i=1}^{N_{gal}}(\zeta_i-\frac{1}{2})}{\sqrt{\sum\limits_{i=1}^{N_{gal}}V_i}}.
\end{equation}

\noindent The test is preformed by computing $T_C$ for sub-samples truncated at increasing apparent magnitude limits.  For complete samples, the values of $T_C$ will follow a Gaussian distribution, with an expectation value of 0 and variance of order unity.  Systematically decreasing negative values of $T_C$ indicate that the sample is becoming incomplete.  We use the $T_C$ statistic to determine the completeness for our main survey volume outside of the zone of avoidance ($|b|>20\degr$) as functions of the $B$-band brightness, 21-cm (HI) flux, and the H$\alpha$+[NII] flux (Figures 1a, 2a, 3a).

By construction, the 11HUGS primary sample has been limited at $B\sim15$ to avoid the severe incompleteness that is known to set in at fainter magnitudes in the original surveys that have provided the bulk of our knowledge about the local volume galaxy population (e.g., Tully 1988b).  To select the sample, we primarily used the NASA Extragalactic Database (NED), with initial cuts based on the Local Group corrected recessional velocity and the ``indicative optical magnitudes'' provided in NED's ``Basic Data'' service.  Direct distance estimates from the literature and information on group membership were then added.  The photometry compilation was refined by adopting measurements from the following large, homogeneous catalogs in the following order of preference (1) $B_T$ from the collection of dwarf galaxy observations obtained by van Zee and collaborators (van Zee 2000, van Zee et al. 1996, van Zee et al. 1997), and Binggeli and collaborators (Barazza et al 2001; Bremnes et al. 1998; Bremnes et al. 2000; Parodi et al. 2002), (2) $B_T$ from the RC3 (de Vaucouleurs et al 1991) as reported on NED, and (3) $B_T$ as compiled and reduced to the RC3 system from HyperLeda.  For 31 galaxies, measurements are not available from one of these sources.  In these cases, the literature was searched and a $B$ magnitude taken from smaller datasets of published photometry.  As a last resort, the ``indicative optical magnitude'' given by NED was adopted in 7 cases.  The resulting collection of data is given in Table 1 of Paper I, and is used to calculate $T_C$ as a function of $B$.  Corrections for Galactic extinction, but not internal extinction, are applied.  The results are plotted in Figure 1a.    $T_C$ drops precipitously below zero for $B\ga15.6$.  This limit of B=15.6 for the 11HUGS main survey volume corresponds to a completeness in $M_B=-14.6$ at 11 Mpc.  In Figure 1b, a plot of $M_B$ with distance is also shown to illustrate the depth of the sample.

We also check the completeness of the 11HUGS sample with respect to the HI gas mass.  To do this, we have compiled 21-cm single-dish fluxes from the literature.  Measurements are available for 96\% of the galaxies in the overall 11HUGS target catalog.  The data are primarily taken from the following three sources, in the following order of preference: the digital archive of Springob et al. (2005) (N=112), the HI Parkes All Sky Survey (HIPASS) catalog as published in Meyer et al. (2004) (N=75), and the homogenized HI compilation of Paturel et al. (2003) as made available through the Hyperleda database (N=215).  Finally, data for 16 galaxies are taken from the HI compilation in the Catalog of Neighboring Galaxies of Karachentsev et al. (2004) and other individual papers.

We evaluate $T_C$ as a function of the HI flux and plot the results in Figure 2a.  Here, the $T_C$ statistic begins to become systematically negative at integrated fluxes $<6$ Jy km s$^{-1}$.  To find the corresponding completeness in the HI mass, we apply the standard relation $M_{HI} [M_{\odot}]\;=\;2.36\;\times\;10^5\;D^2\;F$, where $D$ is the distance in Mpc and $F$ is the 21-cm line flux in Jy km s$^{-1}$.  The gas is assumed to be optically thin, and corrections for the presumably small amount of HI self absorption ($\la$10\%, Haynes \& Giovanelli 1984, Zwaan et al. 2003) are not applied.  The HI mass is plotted against the distance in Figure 2b.  At the edge of the 11 Mpc target volume, a limit of 6 Jy km s$^{-1}$  corresponds to a completeness in $M_{HI}$ down to  2$\times$ 10$^8$ M$_{\odot}$.

Finally, we also compute $T_C$ as a function of the H$\alpha$+[NII] flux and show the results in Figure 3a.  At fluxes below 4$\times$ 10$^{-14}$ ergs s$^{-1}$ cm$^{-2}$, $T_C$ becomes systematically negative.  This corresponds to a completeness of L(H$\alpha$)= 6$\times$  $10^{38}$ ergs s$^{-1}$ at 11 Mpc.  To translate this into a limiting SFR, two issues must be kept in mind.  Most standard conversion recipes (e.g., Kennicutt 1998, which results in a SFR of 0.005 M$_{\odot}$ yr$^{-1}$) are based on expectations for a solar metallicity population, so using this conversion for metal-poor ($\sim$$Z_{\odot}/5$) dwarf galaxies would result in an overestimate of the SFR --- a relative deficiency of metals should cause a greater number of ionizing photons to be produced per unit stellar mass formed (e.g., Lee et al. 2002).  We compute a calibration that is based on a $Z_{\odot}/5$ population using the Bruzual \& Charlot (2003) stellar population synthesis models, assuming Case B recombination and no leakage of Lyman continuum photons from the galaxy.  Relative to a solar metallicity conversion derived from the same models, SFR/L(H$\alpha$) is a factor of $\sim$0.7 lower for the $Z_{\odot}/5$ population.  This SFR scale is shown on the right hand side of Figure 3b.  Another issue is that stochasticity in the formation of high mass ionizing stars may begin to become important for the ultra-low SFRs in this regime.  However, note that a simple calculation, treating the IMF as a probability distribution function (e.g., Oey \& Clarke 2005), shows that such sampling issues become important only at SFRs less than $\sim10^{-4}$ --- a constant SFR of 0.0002 M$_{\odot}$ yr$^{-1}$ over timescales greater than 10 Myr will yield roughly 10 O-stars at any given time, for a Salpeter IMF and mass limits of 0.1 and 1 M$_{\odot}$.

\subsubsection{Luminosity and Mass Functions Comparisons}

Another way of examining the completeness of the sample is to compare the 11HUGS number density distributions to luminosity and mass functions that are based on surveys which have well-determined selection functions (Figures 1c, 2c).  Although we do not expect the densities computed from our small main survey volume ($|b|>20\degr, d<$11 Mpc) to robustly reflect averages taken over much larger patches of the present-day universe (11HUGS poorly samples massive/luminous galaxies while other surveys suffer from large incompleteness corrections in the dwarf galaxy regime) a qualitative comparison of the relative shapes of the distributions still provides an interesting cross-check against the $T_C$ limits computed above. 

In Figure 1c, we plot the Schechter function fits to luminosity functions (LFs) from two independent datasets that have morphological make-ups which should be similar to the 11HUGS sample.  The blue curve is based on the B-band follow-up of the Arecibo HI Strip Survey (AHISS, Zwaan, Briggs \& Sprayberry 2001), a optically blind survey for galaxies selected on 21-cm emission alone.  The red curve is based on spiral and irregular galaxies in the Second Southern Sky Redshift Survey (SSRS2, da Costa et al. 1988; Marzke et al. 1998), which has based its sample on the STScI Guide Star Catalog (Lasker et al. 1990).  The SSRS2 LF shown is a composite of two separate Schechter function fits to the spiral and irregular populations.  Clearly the SSRS2 and AHISS LFs are quite different in their determinations for the densities of dwarf galaxies.  Each of the functions are constrained by fewer than $\sim$30 galaxies for $M_B\gtrsim-15.5$.  This difference emphasizes the uncertainties and illustrates the probable ranges of such measurements for low luminosity populations.  The SSRS2 exhibits a much steeper rise at the faint end, with a slope of $-$1.8, while the slope of the AHISS LF is much flatter at $-$1.0. 

The black curve in Figure 1c show the number densities of 11HUGS galaxies 
in our main survey volume where no internal extinction corrections to the absolute magnitudes have been applied.  The densities based on the 11HUGS sample are systematically higher by a factor of $\sim$2.  This is likely due to cosmic variance.  Since ({\it i}) the characteristic correlation length, the scale on which the density of galaxies exceeds the average by a factor of two, has been well-measured to be $r_{\circ} = 5h^{-1}$ Mpc, ({\it ii}) the power-law slope of the function $\xi = (r_{\circ}/r)^{\gamma}$, which parametrizes the excess probability over random of finding two galaxies separated by a distance $r$, has also been established to be 1.8 (Longair 1998 and references therein), and ({\it iii}) the local volume is not centered on a large void, it is not surprising that the densities we compute are higher relative to those based on surveys which average over much larger volumes.  Karachentsev et al. (2004) also report finding that the integrated luminosity volume density of galaxies within 8 Mpc (based on their Catalog of Neighboring Galaxies) is larger by factors of $\sim$2 than other global measurements.  

To compare the relative shapes at the faint end, the 11HUGS distribution is shifted down by a factor of 2.3 to force approximate agreement near the knee of the LFs (gray dotted curve).  The 11HUGS distributions are not flat at the faint end and show increasing densities for dwarfs that are more consistent with the SSRS2 LF (blue) than the AHISS LF (red).  However, the densities do not continue to rise and abruptly drop for $M_B>-14.5$.  This is consistent with the conclusion based on the $T_C$ statistic that 11HUGS is complete throughout its main survey volume to B=15.6. 

In Figure 2c, we also compare the 11HUGS number densities as a function of the HI mass with HI mass functions (HIMFs) based on the optically blind searches of the Arecibo Dual Beam Survey (ADBS, Rosenberg \& Schneider 2002; blue curve) and HIPASS (Zwaan et al. 2005; red curve).  The Schechter fits for these samples are plotted along with the number densities of the 11HUGS galaxies (black curve).  Here, there is a milder discrepancy between the absolute normalizations.  The gray dotted curves show the 11HUGS distribution shifted down by a factor of 1.4.  Again, Karachentsev et al. (2004) also find that the integrated HI volume density within 8 Mpc is greater by this same factor relative to the HIPASS Bright Galaxy Catalog (Zwaan et al. 2003).  The re-normalized 11HUGS distribution agrees well with the HIPASS HIMF for log($M_{HI}$)$>$8.4 and then drops systematically below it at lower masses.  This is consistent with the $T_C$ statistic assessment that 11HUGS is complete to $M_{HI}=2 \times 10^8  M_{\odot}$.  If the re-normalized 11HUGS distribution is instead compared to the steeper ADBS HIMF, the relative fall-off of our sample begins at a significantly higher log($M_{HI}$) of $\sim$9.  However, this is more likely a reflection of the dependence of the HIMF on environment rather than incompleteness.  Zwaan et al. (2005) show evidence that the HIMF becomes steeper from the field towards higher density regions with $-1.2>\alpha>-1.5$, and the ADBS sample has large contributions from both the Virgo Cluster and the Pisces-Perseus Supercluster.

As for performing this same exercise with our H$\alpha$ luminosity number densities, there is no local H$\alpha$ LF that can be used as a fiducial for comparison since the ones that are available have been based on objective-prism surveys and are highly incomplete for galaxies with modest to low H$\alpha$ equivalent widths (see discussion in Paper I).  Here, we simply plot the 11HUGS L($\alpha$) number densities in Figure 3c.  The distribution sharply falls off below 6$\times$ $10^{38}$ ergs s$^{-1}$, which is the limiting luminosity determined from the $T_C$ statistic.

\section{Starburst Statistics}

\subsection{What is a starburst?}

The goal of this paper is to directly tally the number of currently bursting dwarf galaxies in an approximately volume-limited sample and to compute the fraction of star formation that is concentrated in such systems.  These statistics help provide constraints on the average amplitudes and durations of starburst episodes.  To proceed however, we require criteria that distinguish bursting galaxies from the rest of the population.  This is a non-trivial issue, and many different (and sometimes subjective) criteria have been previously applied (e.g., see overviews given in the proceedings of the 2004 Cambridge Starbursts conference by Gallagher 2005, Heckman 2005 and Kennicutt et al. 2005).  To some extent this variation has been driven by the type of data that are available to gage the star-formation state of the systems under consideration.  For example, while perhaps the most physically fundamental approach is to define starbursts as those galaxies that are forming stars near the maximum possible rate set by causality (i.e., the rate that results when all of the gas in a system is consumed in one dynamical time; Heckman 2005), measurements of the total gas masses and global velocity dispersions that are required to compute the limiting SFRs may not be available, particularly for large samples.  An alternate approach, which is less absolute but can be reasonably applied to present-day galaxies, is to develop a characterization of normal star formation activity based on the overall population of galaxies, and then to identify starbursts as the subset of galaxies which are forming stars at an anomalously prodigious rate with respect to this fiducial (e.g., Gallagher 2005; Kennicutt 2005 and references therein).  We follow such strategies here.

One frequently used definition of a starburst is a galaxy whose current SFR
exceeds its average past value by a factor of two to three (e.g., Hunter \&
Gallagher 1986; Salzer 1989; Gallagher 2005; Brinchmann et al. 2004;
Kennicutt et al. 2005).  The ratio of the current SFR to the past average is
commonly referred to as the stellar birthrate (i.e., the $b$ parameter), and
can be observationally traced by the integrated, galaxy-wide H$\alpha$ EW
(i.e., the H$\alpha$ flux divided by the continuum flux density under the
line).  The integrated EW is thus also related to the specific SFR, the
current SFR normalized by the total mass of stars.  Using the model grids
computed in Kennicutt et al. (1994), and updated in Lee (2006) using the
population synthesis code of Bruzual \& Charlot (2003), an approximate
mapping between $b$ and EW may be constructed.  Assuming a Kennicutt IMF
($\Gamma = -1.5$; Kennicutt 1983), Case B recombination, no leakage of Lyman
continuum photons from the galaxy and no internal extinction, galaxies with
$b>$2-3 should have EW(H$\alpha$) $\gtrsim$ 80-110 \AA.  Thus, one way of
operationally identifying starbursts is by using such thresholds in the
H$\alpha$ EW.

Given the close relationship between the EW and $b$, is it interesting to
study the EW distribution of the galaxies in our sample, and ask whether EW
values of $\sim$80-110 \AA\ coincide with any particular features of the
distribution. In Lee et al. (2007), we initially examined trends in the
$M_B$-EW plane using the measurements given in Table 2, Paper I, and showed
that local star-forming galaxies trace a continuous sequence in EW as well
as morphological type.  The sequence exhibits two characteristic
transitions.  At $M_B\sim-15$ a narrowing of the galaxy locus occurs as the
luminosities increase and morphologies shift from predominantly irregular to
late-type spiral (termed the ``main sequence of star forming galaxies'' by
Noeske et al. 2007).  Above  $M_B\sim-19$, the high-
luminosity end of the sequence turns off toward lower EWs and becomes mostly
populated by intermediate and early-type bulge-prominent spirals.  In the
current analysis, we focus on the low-luminosity galaxies below the upper
transition at $M_B\sim-19$, and divide the galaxies into three equal sized
bins: two in the narrow EW ``waist'' ($-19 \le M_B < -17$ and  $-17 \le M_B
< -15$) and one containing the most extreme dwarf galaxies which lie below
the lower transition where the EW distribution broadens ($-15 \le M_B <
-13$). Histograms of the logarithmic EW in these three bins are shown in
Figure 4.  A histogram of the higher luminosity galaxies ( $-22 \le M_B <
-19$) is also provided for completeness.

To characterize the distributions plotted in Figure 4, we first checked
whether simple Gaussian functions with means and standard deviations
directly computed from the logarithmic EWs in a given luminosity bin would
yield an adequate description of the data.  The gray dashed curves plotted
in each panel represent these functions.  For galaxies with $-19 \le M_B <
-17$, the gray curve provides a good fit, but for the lower luminosity
galaxies (bottom panels), the curves appear somewhat too broad; this is due
to the outliers appearing at the tails of the distributions.  Thus,
to attempt to place more weight on the central components of the distributions, 
Gaussian fits to the histograms minimizing $\chi^2$ were also performed. 
This second set of Gaussian functions is overplotted in yellow, and 
appears to provide an improved fit to the majority of the data with 
$-17 \le M_B < -15$ and $-15 \le M_B < -13$.  This
suggests that one way to describe the distribution of EWs is as a log-normal
function with non-Gaussian excesses at the tails.  In order to check this, we
performed Anderson-Darling tests for normality (Anderson \& Darling 1952) on
both the complete set of logarithmic EWs in each luminosity bin, as well as
sets which were clipped at 3$\sigma$, where $\sigma$ is provided by the fits
to the histograms (see Table 1).  The Anderson-Darling statistic is similar
to the Kolmogorov-Smirnov statistic in that it examines differences between
two cumulative distribution functions to determine if one sample is
consistent with being drawn from the other (the null hypothesis). However,
whereas the K-S test is most sensitive near the median values of the
distribution and does not perform as well at detecting statistically
significant deviations at the ends of distributions, the A-D test is
``stabilized'' by accounting for the variance in the calculated differences
between the distributions and is thus more robust at the tails (Press et al.
1992). Applying the A-D test to our data, we find that the logarithmic EWs
for galaxies with $-19 \le M_B < -17$ are consistent with a Gaussian
function insofar as the null hypothesis cannot be rejected.  As would be
expected from the consistency of the gray and yellow curves, clipping
removes no galaxies from the sample in this bin.  On the other hand, 
for the subsets of EWs in the $-17 \le M_B < -15$ and $-15 \le M_B < -13$ 
bins, the A-D test rejects the null hypothesis at the 99\% and 90\% levels, 
respectively. However, after these subsets are clipped at 3$\sigma$, the null
hypothesis can no longer be rejected with any confidence. We note that
clipping with different thresholds (e.g., 2$\sigma$, 4$\sigma$) yield
distributions that also violate the assumption of normality.
Thus, this exercise provides some support that the central components
of the EWs distributions can be adequately represented by log-normal
distributions (as described by the yellow curves in Figure 4), although
of course, it is by no means proof that the underlying logarithmic
distributions are Gaussian.

What is interesting about this characterization of the EW distributions
is the coincidence between their upper 3$\sigma$ points to the frequently
adopted starburst birthrate thresholds discussed above. 
In the two luminosity bins that are bounded by the EW transitions ($-19 \le
M_B < -17$, $-17 \le M_B < -15$; note that the sample, as discussed in \S\
2, is complete within this regime), the means and dispersions of the yellow
Gaussian curves (Table 1) are essentially the same.  Here,
the EW values of $\sim$80-110 \AA\, which correspond to
starburst-like birthrates of 2-3, occur where the distributions
drop steeply, and the upper value of this range is close to the upper
3$\sigma$ points of the curves (107 \AA\ and 105 \AA\ for $-19 \le M_B <
-17$ and $-17 \le M_B < -15$ respectively). Although our sample is not large
enough to cleanly define the 3$\sigma$ ranges, the correspondence suggests
that for a complete sample of field galaxies with $-19 \le M_B < -15$, the
upper 3$\sigma$ point of the central component of the logarithmic EW
distribution may provide one empirical criterion for identifying dwarf
galaxies currently undergoing global starbursts, at least in the local
Universe.  Such a definition will not nominally predetermine the
starburst number fraction to be 0.135\% (the $>3\sigma$ probability for a
normal distribution) since it appears that a sufficiently populated
distribution should show significant non-Gaussian excesses at the tails,
as has already been discussed.  These excesses are also illustrated in Figure 
5a where the observed logarithmic EW cumulative number distributions are 
compared with those of the best fit Gaussians.   We speculate that the low 
EW outliers are post-burst galaxies.

In the analysis that follows, we compute starburst number and
star formation fractions for a range of EW thresholds which
roughly span stellar birthrates of two to three, using both
fixed EW values as well as those determined by
$\sigma$ limits (Table 1).  In the interest of clarity
in the discussion however, we will primarily adopt results based on an
intermediate EW cut of 100\AA\ which maps to an birthrate value of 2.5. 
As will be shown, the results are robust to variations in the exact choice of 
the threshold between the EW values of $\sim$80-110 \AA\ since, as already 
noted, such values occur where the distributions fall off sharply.
 
\subsection{The Number Fraction of Dwarf Starburst Galaxies}

With the criteria established in the previous section, it is straightforward to determine the number fraction of starbursting dwarf galaxies in the local 11 Mpc volume using data from the H$\alpha$ component of our survey.  In Table 1, we list the fractions and numbers of galaxies with EW$>$80\AA\ ($b\sim2$),  EW$>$100\AA\ ($b\sim2.5$), as well as above and below the 1, 2 and 3 $\sigma$ ranges of EW values in each luminosity bin.  The 3$\sigma$ point for galaxies $-19 < M_B < -15$ maps to $b\sim3$, so the selected EW thresholds roughly span stellar birthrates of two to three.  Cumulative number distributions as a function of the EW are shown in Figure 5a.  Here we quote statistics based on the observed H$\alpha$+[NII] fluxes and EWs which have not been corrected for internal extinction.  Corrections for internal extinction and the contribution of the [NII] lines to the observed flux (as described in Paper I) have an negligible effect on these results.  This is reasonable given that the low luminosity population being considered is metal-poor and has low dust content, and that the [NII] and internal extinction corrections tend to cancel each other out.  Again, we focus the discussion on results based on an
intermediate EW cut of 100\AA, which maps to an birthrate value of 2.5, and then describe the effects of varying the threshold about this value.

Our most robust fractions are based on the $-17 \le M_B < -15$ luminosity bin since this is where the number statistics are tolerable (N = 93), and the sample in our main survey volume ($|b|>20\degr$) is complete (\S\ 2.1).  Dwarf galaxies in this bin should also be outside of the regime where stochasticity in the formation of high mass ionizing stars may affect the interpretation of H$\alpha$ luminosity as a SFR indicator.  As noted at the end of \S\ 2.1.1, such effects may become important for SFR$\lesssim 10^{-3}$ M$_{\odot}$ yr$^{-1}$, but as shown in Figure 6, galaxies with $M_B=-15$ generally have an average SFR of 0.01 M$_{\odot}$ yr$^{-1}$.\footnote{Exactly when IMF sampling issues begin to become relevant is an issue of current debate (e.g., Weidner \& Kroupa 2005; Elmegreen 2006).  Depending on additional assumptions about the cluster mass function, and the relative order in which high and low mass stars form in individual molecular gas clouds, it is claimed that H$\alpha$ could underestimate the SFR by factors $\gtrsim$ 20 beginning at L$_{H\alpha}\sim 10^{39}$ (Pflamm-Altenburg et al. 2007).  Further analysis with the 11HUGS GALEX UV data will be important in constraining these effects since the UV emission probes star formation over a timescale that is 10 times longer than H$\alpha$ (due to its sensitivity to relatively lower mass B-stars), and is less prone to statistical high mass star formation ``flickering.''  A comparative analysis of the 11HUGS H$\alpha$ and FUV SFRs (Lee et al. 2007), as well as independent simulations which probe the impact of Poisson noise on both the UV and H$\alpha$ emission (Tremonti et al. 2007), are underway.  Preliminary results from both studies confirm that SFRs$\gtrsim$ 0.001 M$_{\odot}$ yr$^{-1}$ should be relatively robust to at least these simple stochastic effects.}  Therefore, the subsequent discussion will be based on the statistics from this bin.  Within this luminosity range, systems with EW$>$100\AA\ are rare -- there are only 6 galaxies with EWs this high, and this represents 6\% of the population.  (A listing of these starbursting dwarf galaxies along with some basic properties is given in Table 2, and images are shown in Figure 7.)  The dominant component of the uncertainty in the starburst number fraction is likely due to Poisson noise, so we quote the primary result as 6$^{+4}_{-2}$\%.  The uncertainties given here and below all span the 68\% confidence interval as tabulated in Feldman \& Cousins (1998).  Again, it is important to note that the shape of the distribution ensures that this result is not critically dependent on the exact EW threshold that is adopted, for those values that map to $b\sim2-3$.  If the threshold is dropped down to 80 \AA\ , the results do not change since there are no galaxies between 80\AA\ and 100\AA\ within this luminosity range in the sample.  If the $3\sigma$ point of 105\AA\ is instead used, the results remain the same as well.  Finally, if the threshold is dropped down to 2$\sigma$ above the mean (EW=71\AA), the number count increases to 7 systems (8\%).  

For the higher luminosity galaxies ($-19 \le M_B < -17$), the tails of the EW distribution do not show excesses and there are no global starbursts based on a EW$\gtrsim$100\AA\ cut.  This can partly be attributed to small number statistics, since there are $\sim$35\% fewer galaxies in this bin as compared with the $-17 \le M_B < -15$ bin.  If we assume that the starburst number fraction is also 6\% here, then there should be 4$^{+3}_{-2}$ galaxies with EW$>$100\AA\ for a sample of 59, so it is reasonable that no galaxies with such high EWs are observed.  A Poisson upper limit for the observation of zero events also may be computed, which yields a fraction of $\la 2\%$ in this luminosity regime.  If a EW$>$80\AA\ threshold is instead applied, the fraction is 2\% (N=1).  Dropping further to the upper 2$\sigma$ point (EW=71\AA), the fraction becomes 3\% (N=2).  Comparison of these numbers with the ones computed for the $-17 \le M_B < -15$ bin shows that it is possible that the starburst number fraction decreases for more massive systems. 

If we again apply a threshold value of 100\AA\ to the most extreme dwarfs (for $-15 \le M_B < -13$), the starburst number fractions is 5\% (N=5).   However, the situation is more complicated for this lowest luminosity bin.  First, as discussed in \S\ 2 the completeness steeply drops at $M_B=-14.6$, so statistics reported for this population must be interpreted with caution.  However, it is likely that starburst mass and star-formation fractions computed in this bin represent upper-limits since the low surface brightness dwarfs are more likely to be missing from the sample than the higher surface brightness starbursting systems.  Second, the EW starburst criterion changes because the distribution widens in this regime, so the $3\sigma$ threshold is higher than for the more luminous dwarfs.  This may be partly due to stochastic sampling of the upper part of the IMF as discussed above, and/or burstier star formation in in the lowest mass systems.  If the 2$\sigma$ and 3$\sigma$ values of this bin are instead used (i.e., 115\AA\ and 242\AA), the number fractions instead are 5\% and 3\% respectively.  

Finally, the results from all three bins (for $EW>100$\AA) may be compared.  Beginning with the most luminous bin, the starburst number fractions are  $\lesssim$2\%, 6$^{+4}_{-2}$\% and 5$^{+5}_{-3}$\%.  The local starburst number fraction may thus only be a weak function of luminosity. 

\subsection{The Fraction of Star Formation Occurring in Starbursting Dwarfs}
  
Arguably however, a more important statistic is the fraction of star formation that occurs in starburst episodes.  If the star formation fraction is high, this could be evidence that the burst mode dominates the evolution of dwarfs, even if the number fraction is low.  This statistic is also more robustly measured, as it is relatively immune to incompleteness in the less active galaxies that are most likely to be missed, as such systems would not add considerably to the H$\alpha$ volume density.

To estimate the star formation fraction, we sum the observed H$\alpha$+[NII] luminosities ($L_{H\alpha+[NII]}$) as a function of the EW.  Again, applying corrections for internal extinction and the contribution of the [NII] lines to the observed flux as described in Paper I does not significantly impact the results.\footnote{Dust corrections for individual objects in the sample may also be computed by using the Spitzer MIPS imaging data that is being obtained by the Local Volume Legacy Survey (see \S\ 2).  Consistency between burst statistics incorporating such corrections and those reported here should be checked in future work, particularly for the higher luminosity dwarfs ($M_B\lesssim-17$) .}  The percentage of $L_{H\alpha+[NII]}$ contained in galaxies with EW$>$100\AA\ ($b\sim2.5$), EW$>$80\AA\ ($b\sim2$), as well as above and below the 1, 2 and 3 $\sigma$ ranges of the EW are reported in Table 1, and the cumulative $L_{H\alpha+[NII]}$ distribution as a function of the EW are plotted in Figure 5b.  

For galaxies with $-17 \le M_B < -15$, the six systems with EW$>$100\AA\ are responsible for 23\% of all of the star formation occurring in this luminosity bin.  A simple translation of the Poisson error in the number fraction results in an error of $^{+14}_{-9}$\% in the $L_{\alpha}$ fraction.  From Figure 5b, it can be seen that the starburst system that has the largest SFR contributes $\sim$10\% (NGC 3125), so our simple estimation of the error appears to be reasonable.  This quantity is also not highly sensitive to the exact EW value used to classify starbursts.  A lower EW$>$80\AA\ cut or a slightly higher 3$\sigma$ point of 107\AA\ does not change the result, while applying a lower 2$\sigma$ threshold only causes a small rise of the fraction to 25$^{+16}_{-8}$\%.  

For $-19 \le M_B < -17$, no systems are observed with EW$>$100\AA\ as described in the previous section.  However, the 2 galaxies with EWs higher than the 2$\sigma$ threshold of 71\AA\ produce 9\% of the H$\alpha$ luminosity, and this is consistent with the ratio of 7 galaxies with $-17 \le M_B < -15$ and EW$>$71\AA\ producing 25\% of the H$\alpha$ luminosity in that bin.  Therefore, estimating that each starburst galaxy is on average responsible for $\sim$4\% of the total star formation occurring in the population, we infer that the fraction of star formation due to EW$>$100\AA\ galaxies in this luminosity range must be less than $\sim$5\%.  Application of an EW$>$80\AA\ cut yields star formation fraction of 4\%, and a further drop to the 2$\sigma$ point results in 9\%.

Finally, the fraction of star formation occurring in the lowest luminosity starbursts is computed, keeping in mind the caveats on incompleteness and stochasiticity in the formation of ionizing stars that affect the sample in this bin.
Systems exceeding EWs of 100\AA\ are responsible for 16\% of the star formation.  Dropping the cut to an EW of 80\AA\ results in a small increases of starburst star formation fraction to 20\%.
If the $\sigma$ thresholds defined by the broader EW distribution for $-15 \le M_B < -13$ are instead used, galaxies with EW greater than 115\AA\ (2$\sigma$) and 242\AA\ (3$\sigma$) are responsible for 16\% and 13\% of the star formation respectively.

From these calculations, it is clear that a significant amount of the overall star formation in present-day dwarf galaxies takes place in starbursts. However, the results also clearly show that a more continuous, steady state of star formation dominates in the present epoch, both in terms of being the mode that operates during the vast majority of the time and in which most of the stars are being created.  The average burst amplitude does not appear to vary strongly with luminosity, with 4\% of the total star formation density in a given luminosity bin being concentrated in individual starbursting systems.

\section{Discussion}
\subsection{Comparison with Previous Work}

Based on the 11HUGS sample, 6$^{+4}_{-2}$\% of late-type dwarf galaxies are starbursting, and 23$^{+14}_{-9}$\%  of the overall star formation in dwarfs occurs in the starburst mode.   The numbers are quoted from the luminosity bin where the sample is both complete and has tolerable number statistics (for $-17 \le M_B < -15$), but the flanking bins (for $-19 \le M_B < -17$ and $-15 \le M_B < -13$) do appear to have fractions that are consistent within the Poisson errors.  In this section we compare these results with previous estimates.

With respect to the starburst number fraction, there is good consistency between our result and previous estimates, which is notable because the studies cover a range of independent approaches to the problem, and use different starburst selection criteria.  Early work compared the space densities of low luminosity UV-selected Markarian galaxies to those of field galaxies (Sargent 1972, Huchra 1977) in order to constrain the fraction of ``flashing'' systems.  These studies produced tentative estimates of 7\% at $M_{p}=-$17, and 10\% at $M_{p}=-$14, based on very small samples containing about a dozen objects each, and large incompleteness corrections.  Later, Salzer (1989) also examined the space densities of active galaxies, using the University of Michigan [OIII]$\lambda$ 5007 emission-line selected survey.  The comparison set of normal galaxies was derived from the magnitude-selected Catalog of Galaxies and Clusters of Galaxies (Zwicky 1961-1968) and the UGC (Nilson 1973).  The number statistics of the Salzer (1989) samples were improved relative to the earlier Markarian studies, with $N\sim40$ and $N\sim30$ in the emission-line selected and general population samples respectively.  Using the numbers reported there, we calculate that the low-luminosity Michigan emission-line galaxies ($-17\leq M_B < -15$) represent $\sim$10\% of the general population in that luminosity range.

Kauffmann et al. (2003a,b) have used an entirely different method to calculate the fraction of bursty galaxies as a function of stellar mass in the Sloan Digital Sky Survey (SDSS).  In this work, an extensive grid of stellar population synthesis models, spanning a range of metallicities and star formation histories, is constructed.  A Bayesian technique is then used to compare the observed and modeled H$\delta$ absorption and 4000 \AA\ break to generate a likelihood distribution of the fraction of stellar mass formed in bursts for each galaxy in their sample.  Two statistics are reported: (i) ``$F_{burst}(50\;\mbox{per cent})>0$,'' the fraction of galaxies whose likelihood distribution of burst masses have {\it median} values greater than zero, and (ii) ``$F_{burst}(2.5\;\mbox{per cent})>0$,'' the fraction of ``high confidence bursty galaxies'' whose likelihood distributions have their lower 2.5 percentile point above zero.  For galaxies with 8.0$<$log$M_{\ast}<$8.5 (which corresponds to $-16\la M_B \la -14.5$ assuming an approximate $M/L_{B}=$1 which is typical for dIrr galaxies; e.g. Miller \& Hodge 1994; Lee 2006), Kauffmann et al. find that the fraction with $F_{burst}(50\;\mbox{per cent})>0$ is over 50\%, while $F_{burst}(2.5\;\mbox{per cent})>0$ is 9\%.  For higher mass dwarfs with 8.5$<$log$M_{\ast}<$9.0 ($-17\la M_B \la -16$), the values instead are 36\% and 4\%, respectively.  Clearly, the number fractions estimated using $F_{burst}(50\;\mbox{per cent})>0$ to discriminate whether a galaxy has undergone a burst in the recent past are too large to be consistent with any of the other estimates discussed above.  This is perhaps not too surprising since half of the models which are consistent with the observations for the $F_{burst}(50\;\mbox{per cent})>0$ galaxies have SFHs in which there have not been any bursts at all, and the burst number fractions computed in this way are probably overestimates.  However, the fractions of ``high confidence bursty galaxies'' are in good agreement with the other measurements.  Thus, it appears that the starburst number fraction for dwarf galaxies with $M_B\ga-15$ is well determined, and moreover, relatively robust to the method used to pick out starbursts.  Past estimates consistently lie between 4\% and 10\%, and our 11HUGS measurement of 6$^{+4}_{-2}$\% is representative of these values.

Kauffmann et al. (2003a,b) also have reported a decrease of the burst number fraction with increasing stellar mass.
When their $F_{burst}(50\;\mbox{per cent})>0$ criterion is used to identify galaxies with recent bursts, the number fraction of starbursts drops from 54\% at $\sim$10$^8$ M$_{\odot}$ to 12\% $\sim10^{10}$M$_{\odot}$.  However, when the analysis is restricted to the ``high confidence bursty galaxies'' the absolute decline is much smaller, changing from 9\% to 1\%.  The latter result is more consistent with our finding that the number fraction is $\lesssim$10\% over 7 magnitudes in $M_B$.

As for the fraction of star formation which takes place in bursting systems, there appears to be few prior studies which have attempted to place constraints on this quantity for dwarf galaxies per se.  Based on the SDSS, Brinchmann et al. (2003) have found that starbursts (which they define as galaxies with $b\geq2-3$) are responsible for 20\% of the local star formation density.  Although this is consistent with our 11HUGS estimate of 23$^{+14}_{-9}$\% for dwarf galaxies, the Brinchmann measurement refers to the total galaxy population.  

As another check on the starburst star formation fraction, we perform a coarse calculation using the KPNO International Spectroscopic Survey (KISS; Salzer et al. 2000), a 2nd generation CCD-based, emission-line selected, objective-prism survey.  We use the sample of H$\alpha$- selected galaxies cataloged in List 1 (Salzer et al. 2001), which contains 1128 candidates identified from the objective-prism images.  Follow-up slit spectroscopy has been completed for all of these candidates and 907 of them are classified as star-forming galaxies based on their [OIII]$\lambda$5007/H$\beta$ and [NII]$\lambda$6584/H$\alpha$ line ratios.  Completeness of the sample has been assessed through the standard $V/V_{max}$ test (Schmidt 1968), and limiting volumes have been calculated for each object (Salzer, private communication; also see Lee et al. 2002).   We use these volumes to calculate star formation rate densities as a function of the H$\alpha$ EW.   The H$\alpha$ fluxes and EWs that are measured from the objective-prism spectra are used in this calculation, since they are less likely to suffer from aperture effects and will be closer to the integrated values than those measured from the slit spectra.  For $-17 < M_B < -15$ (N=61), we find that KISS galaxies with EW$>$100\AA\ (N=21) and EW$>$70\AA\ (N=34) are responsible for 22\% and 38\% of the total $L_{H\alpha}$ output in this bin.  We also can compute the starburst number fractions comparing the number densities of high-EW KISS galaxies with those from a sample which better represents the overall population of dIrrs.  Using the B-band luminosity function determined by Zwaan et al. (2001), which is based on the HI selected sample from the Arecibo HI Strip Survey (Zwaan et al. 1997) to provide the denominator of the fraction, we find that galaxies with EW$>$100\AA\ comprise 5\% of the dwarf population, while it is 12\% for the lower threshold of EW$>$70\AA.  These results, which are based on better number statistics for the high EW galaxies, are in good agreement with the fractions estimated from 11HUGs and elsewhere.

\subsection{Constraints on the Dwarf Galaxy Starburst Duty Cycle and Associated Uncertainties}

Our star formation census within 11 Mpc robustly shows that dwarfs that are 
currently experiencing massive global bursts (as identified
as galaxies with integrated H$\alpha$ EWs exceeding 100 \AA) are just the 
$\sim$6\% tip of a low-mass galaxy iceberg, and that the bulk ($\sim$70\%) 
of star formation in the overall population takes place in a more continuous mode.
To make more detailed inferences regarding the characteristic values of burst parameters, 
some assumptions must be made.


The starburst number and star formation fractions of a volume-limited galaxy sample provide constraints on the duty cycle since the relative number of galaxies observed in various phases of the cycle will scale with the relative durations of those phases.  The statistics can be interpreted most directly in the simplest scenario where (1) there are only two phases, an ``on'' (burst) phase and an ``off'' (quiescent) phase and (2) the objects share a common average star formation history such that bursts can occur with equal probability in all galaxies on which the statistic is based.  Under such ``equal probability'' assumptions, the fraction of star formation due to starbursts is equivalent to the fraction of stellar mass formed in the burst mode, the starburst number fraction is equivalent to the fraction of time spent in the burst mode, and the ratio of the two reflects the average amplitude during the burst.  Based on the 11HUGS sample then, the duty cycle is 6$^{+4}_{-2}$\%, the fraction of stars formed in the bursts is 23$^{+14}_{-9}$\%, and the SFR in the burst mode is on average $\sim$4 times greater than in the quiescent mode.  With the additional assumption that the typical duration of a global starburst is $\sim 100$ Myrs (based on the CMD-based SFHs of N1569, Vallenari \& Bomans 1996; Sextans A, Dohm-Palmer et al. 2002; DDO 165, Weisz et al., 2008), such events can roughly be estimated to occur every 1-2 Gyr.  If the equal probability assumptions are incorrect, and instead the starburst mode only operates in a particular sub-set of the population, then the statistics represent lower limits on the duty cycle and frequency.

\subsubsection{Non-detections --- Inferences Regarding the Inter-burst Phase}
As mentioned in Paper I, spiral and irregular galaxies which are completely devoid of recent star formation appear to be rare.  This has also been noted by Meurer et al. (2006) and James et al. (2008) who have carried out complementary H$\alpha$ imaging surveys of nearby galaxies (SINGG, Meurer et al. 2006, and the H$\alpha$GS, James et al. 2004).  In 11HUGS, H$\alpha$ emission is undetected in only 22 out of 410 galaxies which have been observed.  All of these galaxies are in our secondary sample (with $B>15$ or $|b|<20\degr$ or T$\leq$0), and were added to the dataset as observing time allowed or as available from the literature.  Three galaxies, NGC 5206, UGC 2689 and NGC 3115, have S0 morphologies, so a lack of H$\alpha$ emission is not surprising.  The characteristics of the non-detections (in our main survey volume, $|b|>$20\degr; N=20) are illustrated in Figures 1b, 2b, and 3b (green symbols), where $M_B$, $M_{HI}$ and $L_{H\alpha+[NII]}$ are plotted as a function of distance\footnote{In Figure 1b, two galaxies (LGS3 and Leo T) are fainter than the lower bound of the plot.  In Figure 2b, one galaxy (BK3N) does not have a published 21-cm flux and is not shown.}.  The L$_{H\alpha}$ upper-limits shown in Figure 3b correspond to 5$\sigma$ point source detections.

Of the 19 H$\alpha$ non-detections with late-type morphologies, all are extremely low-luminosity galaxies with $-13.6 \ge M_B \ge -7.9$.  HI measurements are available for most of these latter systems, and show that they uniformly have very low gas masses ($\lesssim 10^8$ M$_{\odot}$).  Two of these, LGS3 and Leo T are ``transition''-type systems (e.g., Grebel et al. 2003), with morphologies between that of dIs and dE/dS0s.  Thus, the lack of H$\alpha$ emission either reflects the lack of sufficient fuel for star formation, or SFRs that are so low that the probability of forming a high mass ionizing star and/or of observing an HII region at any given time are small. 

The H$\alpha$ luminosities are $<$10$^{37}$ ergs s$^{-1}$, which is in the regime of nebular photoionization by single O stars (e.g., Oey \& Kennicutt 1997).  Thus, Poisson fluctuations can result in an absence of O-stars (and hence H$\alpha$ emission), although lower mass star formation may still be occurring.  Recent and/or on-going star formation in the majority of H$\alpha$ non-detections in our sample therefore cannot be ruled out.

The fact that H$\alpha$ emission is observed in 95\% of the galaxies in the 11HUGS sample may not be too surprising, however, given the general morphological restriction to spiral and irregular galaxies in our original sample selection.  After all, spirals are classified as such precisely because of the presence of star formation due to spiral density waves; i.e. visible spiral patterns are due concentrations of young stars and HII regions and not concentrations of faded older populations unaccompanied by recent star formation.  The same can thus be said for the dwarf irregulars, whose lumpy structure must be indicative of current star-formation.  {\it The interesting implication however is that if all of the dwarfs in this sample are equally prone to bursting, as in the simple ``equal probability'' scenario above, and the inter-burst phase is represented by the normal galaxies in the sample, then `off' modes rarely occur and the inter-burst state must be characterized by lower-levels of star formation rather than by its complete cessation.}   This is consistent with the CMD analysis of the starburst dwarf galaxy N1705 by Annibali et al. (2003), who find no evidence for $\sim$100 Myr gaps in its star formation over the past Gyr, and assert that this is a result common to all other blue compact dwarfs with resolved stellar population studies.  Under these assumptions, the characteristic SFR in the inter-burst state would be given by the average relationship between the SFR and $M_B$ (Figure 6). 

Characterization of the star formation activity in the inter-burst phase also enables the post-burst luminosity fading to be constrained.  In the equal probability scenario, normal galaxies have an average birthrate of 0.5 (EW$\sim$30\AA) and burst amplitude of $\sim$4.  Using the same models described in \S\ 3.1, this would result in a modest factor of two fading.  The most robust approach however uses SFHs derived from the color magnitude diagrams of resolved stars, as it obviates the need for assumptions about the evolutionary relationship between sub-classes of objects that we have made here.  Detailed comparisons between such SFHs and the statistically inferred one reported here should be the subject of future work.

\subsubsection{The ``Equal Probability'' Assumption}
Whether dwarfs with sufficient gas reservoirs generally cycle between similar quiescent and global starburst phases has been the subject of a long-running debate (e.g., Gallagher et al. 1984; Marlowe et al. 1999; Simpson \& Gottesman 2000; van Zee 2001; Hunter \& Elmegreen 2004; Pelupessy et al. 2004; Gil de Paz \& Madore 2005; Bekki 2008).  

This unresolved issue is the greatest source of uncertainty in our constraints on the duty cycle (i.e.; uncertainty in the denominator of the fractions).  While we do not conclusively address the problem here, insight can be gained by examining the properties of the galaxies picked out by the starburst criterion established in \S\ 3.1 within the context of previous work.  

From the observational side, the arguments have generally been against a common burst mode and have mainly come in two flavors.  The first is an application of Occam's razor, and argues that since the properties (e.g., integrated $UBV$ colors, H$\alpha$ stellar birthrates and gas-depletion timescales) of the majority of dwarf irregulars can be explained by a simple constant SFR over the galaxies' lifetime, bursts need not be invoked and are probably not a common phase in the star formation histories of these systems (e.g., Gallagher et al. 1984; van Zee 2001).  It is therefore posited that only a small sub-class of dwarf irregulars undergo bursts.  The second argument is based on the comparative structural parameters of normal dwarf irregulars and starbursting systems.  The surface brightness profiles of starbursting dwarfs have been shown to be described by two components, an exponential outer envelope, and a blue central excess where the bulk of the star formation is concentrated (e.g., Papaderos et al. 1996, Marlowe et al. 1999, Gil de Paz \& Madore 2005).  The underlying envelopes of starbursts on average have higher central surface brightnesses and shorter scale lengths than normal dwarf irregulars.  Based upon these structural differences, such studies argue that a general evolutionary connection between the two types of systems is unlikely.  Rather, they argue, only the highest surface brightness, most compact dwarf irregulars are thought to host such events.

In Table 2, we list the dwarf galaxies with the highest EWs ($2\sigma$ above the mean logarithmic EW as given in Table 1) in both the 11HUGS main ($|b|>20\degr$) and secondary ($|b<20\degr$) survey volumes.  In Figure 7, emission-line only H$\alpha$+[NII] and $R$-band images are shown for the six galaxies with EWs exceeding 3$\sigma$ of the logarithmic mean in our best populated and complete luminosity bin ($-17 < M_B < -15$).  The objects picked out by our starburst criterion are generally well-studied (given that they are nearest such systems), and many have been examined in previous structural studies (e.g., N1705, N3125, N5253, Marlowe et al. 1999; N1705, ESO435-IG020, Gil de Paz \& Madore 2005).  They are typical of the compact, high surface brightness cored objects embedded in exponential envelopes.  Thus, if the conclusions of the structural studies are correct and only the dwarf irregulars with similarly compact exponential profiles undergo bursts, then the duty cycle may be much larger than the 6\% calculated here.  Follow-up analysis of the broad-band structural parameters of the 11HUGS complete sample may thus provide more insight into the statistics of the progenitor population and alternate constraints on characteristic burst cycle properties.  Such constraints are likely to be upper-limits, and thus when combined with the statistics reported here may reasonably bound the range of true values.

\section{Conclusions}

In this paper we have used the H$\alpha$ component of the 11 Mpc H$\alpha$ UV Galaxy Survey (11HUGS) to quantify the prevalence of global starbursts in dwarf galaxies in the present day universe, and to infer their characteristic duty cycles and amplitudes.  A summary of our findings is as follows:  
       
\begin{enumerate}    

\item The galaxy sample within the 11HUGS main survey volume ($|b|>20\degr$, d$ <$11 Mpc) is found to be complete to $M_B<-14.6$ and $M_{HI}>2\times10^8 M_{\odot}$ using the $T_c$ statistic of Rauzy (2001).  Ancillary checks involving the comparison of 11HUGS luminosity densities to independently derived luminosity functions are consistent with these limits.  

\item To identify dwarf galaxies currently undergoing global starbursts, 
we use an integrated H$\alpha$ EW threshold of 100\AA, which correponds 
to a stellar birthrate of $\sim$2.5, and also explore empirical starburst
definitions based on $\sigma$-thresholds of the observed
logarithmic EW distribution.  The distribution of integrated H$\alpha$ equivalent widths (EWs) can be characterized by log-normal functions that show non-Gaussian excesses at the tails.  For galaxies with $-19 \le M_B < -15$, the correspondence of the upper 3$\sigma$ point of the central components of these distributions to birthrate parameter values around the starburst threshold ($b\sim$3) suggests that dwarf galaxies in the local Universe currently undergoing global starbursts may be identified as outliers at the high EW end. 
We speculate that the low EW outliers are post-burst galaxies.

\item The dwarf galaxy starburst number fraction is shown to be 6$^{+4}_{-2}$\% while the fraction of stars formed in such systems is 23$^{+14}_{-9}$\%.  These results are primarily based on (i) galaxies with $-17 \le M_B < -15$, which is the luminosity bin that is the most robustly populated and statistically complete in the present sample and (ii) a definition which identifies dwarf stabursts as those systems with integrated H$\alpha$ EW exceeding 100\AA.  From these statistics we conclude that a continuous, steady state of star formation dominates in the present epoch, both in terms of being the mode that operates during the vast majority of the time and in which most of the stars are being created.  These results are not sensitive to modest changes in the EW threshold used to define starbursts.  There is good consistency between our results and previous estimates, which is notable because the studies cover a range of approaches to the problem and use different starburst selection criteria.

\item Spiral and irregular galaxies which are devoid of recent star formation appear to be rare.  The H$\alpha$ non-dectection rate for our sample is $\sim$5\%.  The majority of the undetected galaxies are extremely low-luminosity irregulars ($M_B<-13.6$), where lack of H$\alpha$ emission may not necessarily indicate a lack of current star formation, because of stochastic effects in high mass star production in galaxies where the lifetime averaged SFRs are on the order of $10^{-4}$ M$_{\odot}$ yr$^{-1}$.  An inference is that `off' modes rarely occur in starburst cycles and that the inter-burst state must be characterized by low-levels of star formation rather than by its complete cessation, at least for dwarfs with $M_B\lesssim-15$.

\item  In the simplest scenario where the dwarf galaxies in our sample all share a common average star formation history such that bursts can occur with equal probability in every system, (i) the 6$^{+4}_{-2}$\% dwarf galaxy starburst number fraction can be directly interpreted as the duty cycle (the fraction of time spent in the burst state), (ii) the fraction of stars formed in the burst mode is 23$^{+14}_{-9}$\% and (iii) SFR in the burst mode is on average $\sim$4 times greater than in the quiescent mode.  If the assumptions are incorrect, and instead the starburst mode only operates in a particular sub-set of the population, then the statistics represent lower limits on the duty cycle.  Whether this assumption holds has been the subject of the long-running debate in studies of dwarf galaxy, but future work on the resolved stellar populations of larger samples of dwarf irregulars, in conjuction with the analysis presented here, have the potential of clarifying this issue.

\end{enumerate}

\acknowledgments

We are grateful to Evan Skillman, Don Garnett and Deidre Hunter for valuable feedback on earlier versions of this work.  We thank the anonymous referee for useful comments and suggestions which have helped improved the clarity of this paper.
Helpful conversations with Liese van Zee, Luis Ho, Pat McCarthy and Alan Dressler are also acknowledged.  This research has made use of the NASA/IPAC Extragalactic Database (NED) which is operated by the Jet Propulsion Laboratory, California Institute of Technology, under contract with the National Aeronautics and Space Administration.  Use of the HyperLeda database (http://leda.univ-lyon1.fr) is also acknowledged.

\clearpage



\clearpage

\begin{table}[p]
\caption{Starburst Number and Star Formation Fractions}
\footnotesize
\begin{tabular}{rcccccc}
\\
\hline\hline
&$-19.0\leq M_B<-17.0$ &$-17.0\leq M_B<-15.0$ &$-15.0\leq M_B<-13.0$\\
\multicolumn{1}{c}{\small Quantity}&($N=59$)&($N=93$)&($N=104$)\\
\hline
N(EW$>100$\AA) &0&6&5\\
\%(EW$>100$\AA)&0\%&6\%&5\%\\
\%$L_{H\alpha}$(EW$>100$\AA)&0\%&23\%&16\%\\
\hline
N(EW$>80$\AA)&1&6&8\\
\%(EW$>80$\AA)&2\%&6&8\%\\
\%$L_{H\alpha}$(EW$>80$\AA)&4\%&23\%&20\%\\
\hline
&\multicolumn{3}{c}{Logarithmic H$\alpha$ Distribution Statistics}\\

&$<$lg(EW)$>$=31\AA&$<$lg(EW)$>$=32\AA&$<$lg(EW)$>$=26\AA\\
\hline
1$\sigma$ range&20\AA, 47\AA&22\AA, 48\AA&12\AA, 55\AA\\
$N\leq1\sigma$, $N\geq1\sigma$&12, 11&21, 19&26, 17\\
\%$\leq1\sigma$, \%$\geq1\sigma$& 20\%, 19\%&23\%, 20\%&24\%, 16\%\\
\%$L_{H\alpha}\leq1\sigma$, \%$L_{H\alpha}\geq1\sigma$&11\%, 33\%&10\%, 41\%& 5\%, 32\% \\
\hline
2$\sigma$ range&13\AA, 71\AA& 15\AA, 71\AA& 6\AA, 115\AA\\
$N\leq2\sigma$, $N\geq2\sigma$&2,2 &13, 7&9, 5\\
\%$\leq2\sigma$, \%$\geq2\sigma$&3\%,3\%& 14\%, 8\%& 8\% 5\%\\
\%$L_{H\alpha}\leq2\sigma$, \%$L_{H\alpha}\geq2\sigma$&$<$1\%, 9\%& 4\%, 25\%& $<$1\%, 16\%\\
\hline
3$\sigma$ range&9\AA, 107\AA& 10\AA, 105\AA&3\AA, 242\AA\\
$N\leq3\sigma$, $N\geq3\sigma$&0, 0&8, 6& 5, 3\\
\%$\leq3\sigma$, \%$\geq3\sigma$&0\%, 0\%& 9\%, 6\%& 5\%, 3\%\\
\%$L_{H\alpha}\leq3\sigma$, \%$L_{H\alpha}\geq3\sigma$&0\%, 0\%&2\%, 23\% & $<$1\%, 13\%\\
\hline\hline
\end{tabular}
\end{table}

\clearpage

\begin{deluxetable}{lrrrrcclrcllc}
\tabletypesize{\scriptsize}
\rotate
\tablecolumns{13}
\tablewidth{0pc}
\tablecaption{11HUGS Dwarf Galaxies with EWs Exceeding 2$\sigma$ of the Logarithmic Mean}

\tablehead{

\colhead{Galaxy Name}   &
\colhead{RA}            &
\colhead{DEC}           &
\colhead{$b$}           &
\colhead{$cz$}          &
\colhead{D}             &
\colhead{method}        &
\colhead{$M_B$}           &
\multicolumn{3}{c}{EW(H$\alpha$+[NII])}  &
\colhead{SFR}           &
\colhead{M(HI)}		
\\

\colhead{}           &  
\colhead{[J2000]}           &
\colhead{[J2000]}           &
\colhead{[degrees]}           &
\colhead{[km s$^{-1}$]}           &
\colhead{[Mpc]}           &
\colhead{}           &
\colhead{[mag]}           &
\multicolumn{3}{c}{[\AA]}  &
\colhead{[M$_{\odot}$ yr$^{-1}$]}          &
\colhead{[M$_{\odot}$]}          
\\

\colhead{(1) }           &  
\colhead{(2) }           &
\colhead{(3) }           &
\colhead{(4) }           &
\colhead{(5) }           &
\colhead{(6) }           &
\colhead{(7) }           &
\colhead{(8) }           &
\multicolumn{3}{c}{(9)}  &
\colhead{(10) }          &
\colhead{(11) }          
}
\startdata
\multicolumn{13}{c}{$-19 \le M_B < -17$; $|b|>20\degr$}\\ 
\hline
NGC4449	     &122811.2	&440536	  &72.40   &207	 &4.21	&trgb  &-18.2 &72 &$\pm$  &5   &0.65  &1.0e9\\
NGC4656	     &124357.7	&321005	  &84.70   &646	 &8.6	&v(LG) &-18.8&96 &$\pm$  &9    &1.0   &5.2e9\\
\hline
\multicolumn{13}{c}{$-17 \le M_B < -15$; $|b|>20\degr$}\\ 
\hline
ESO409-IG015 & 000531.8 &-280553  & -79.79 &737  &10.4	&v(LG) &-15.0 &262 &$\pm$ &26  &0.053 &1.8e8\\
NGC1705      & 045413.7 &-532141  & -38.74 &633  &5.1	&trgb  &-15.9 &109 &$\pm$ &7   &0.093 &9.1e7\\
NGC2366      & 072854.6 &691257   & 28.53  &80   &3.19	&trgb  &-16.3 &149 &$\pm$ &38  &0.13  &6.3e8\\
ESO435-IG020 & 095920.7 &280754   & 21.03  &971  &9.0	&v(LG) &-15.7 &174 &$\pm$ &21  &0.12  &1.5e8\\
NGC3125      & 100633.6 &-295609  & 20.64  &1113 &10.8	&v(LG) &-17.0 &221 &$\pm$ &25  &0.51  &2.6e8\\
NGC4485	     & 123031.1	&414201	  & 74.81  &493	 &7.1	&v(LG) &-17.0 &76  &$\pm$ &13  &0.14  &3.4e8\\
NGC5253      & 133955.9 &-313824  & 30.10  &407  &3.15	&ceph  &-16.8 &120 &$\pm$ &9   &0.23  &1.3e8\\
\hline
\multicolumn{13}{c}{$-15 \le M_B < -13$; $|b|>20\degr$}\\ 
\hline
 UGC4459	&083407.2	&661054	   &34.95	&20	&3.56	&trgb	  &-13.1 &125 &$\pm$	&54 &0.0079 &4.1e7 \\
 MRK36	        &110458.5	&290822	   &66.49	&646	&7.8	&v(flow)  &-13.8 &467 &$\pm$	&33 &0.042  &1.5e7\\
 UGCA281	&122616.0	&482937	   &68.08	&281	&5.7	&bs	  &-13.5 &335 &$\pm$	&17 &0.044  &6.7e7\\
 MRK475	        &143905.4	&364821	   &65.31	&583	&9.0	&v(flow)  &-14.3 &314 &$\pm$	&16 &0.027  &2.9e6\\
 ESO140-G019	&182246.4	&-621613   &-20.61	&950	&10.8	&v(flow)  &-13.7 &135 &$\pm$	&20 &0.020  &1.4e8\\
\hline
\multicolumn{13}{c}{$-19 \le M_B < -13$; $|b| <20\degr$ (secondary sample)}\\
\hline
NGC1569	     &043049.0	&645053  &11.24	   &-104 &1.9	&bs     &-17.1 &215 &$\pm$ &12 &0.43  &7.2e7\\
UGCA116	     &055542.6	&032330	 &-10.77   &789	 &9.1	&v(LG)	&-17.2 &451 &$\pm$ &23 &1.13  &3.4e8\\
ESO495-G021  &083615.4	&-262434 &8.58	   &873	 &7.8	&v(LG)	&-17.5 &134 &$\pm$ &8  &0.70  &1.9e8\\
NGC2835	     &091752.9	&-222118 &18.51	   &886	 &8.0	&v(LG)	&-18.9 &88  &$\pm$ &10 &1.29  &2.0e9\\
NGC5408	     &140321.0	&-412244 &19.50	   &506	 &4.81	&trgb	&-16.5 &121 &$\pm$ &12 &0.14  &3.4e8\\
ESO137-G018  &162059.2	&-602916 &-7.43	   &605	 &6.4	&trgb	&-17.9 &94  &$\pm$ &26 &0.16  &3.5e8\\
IC4662	     &174706.4	&-643825 &-17.85   &302	 &2.44	&trgb	&-15.5 &101 &$\pm$ &10 &0.078 &1.8e8\\
\enddata
\tablerefs{Columns 1---9 as in Tables 1 \& 3 in Paper I.  Columns 10---11: See \S\ 2.1.1.} 

\end{deluxetable}


\clearpage

\thispagestyle{empty}
\setlength{\voffset}{-20mm}
\begin{figure}[p]
\epsscale{0.4}
\plotone{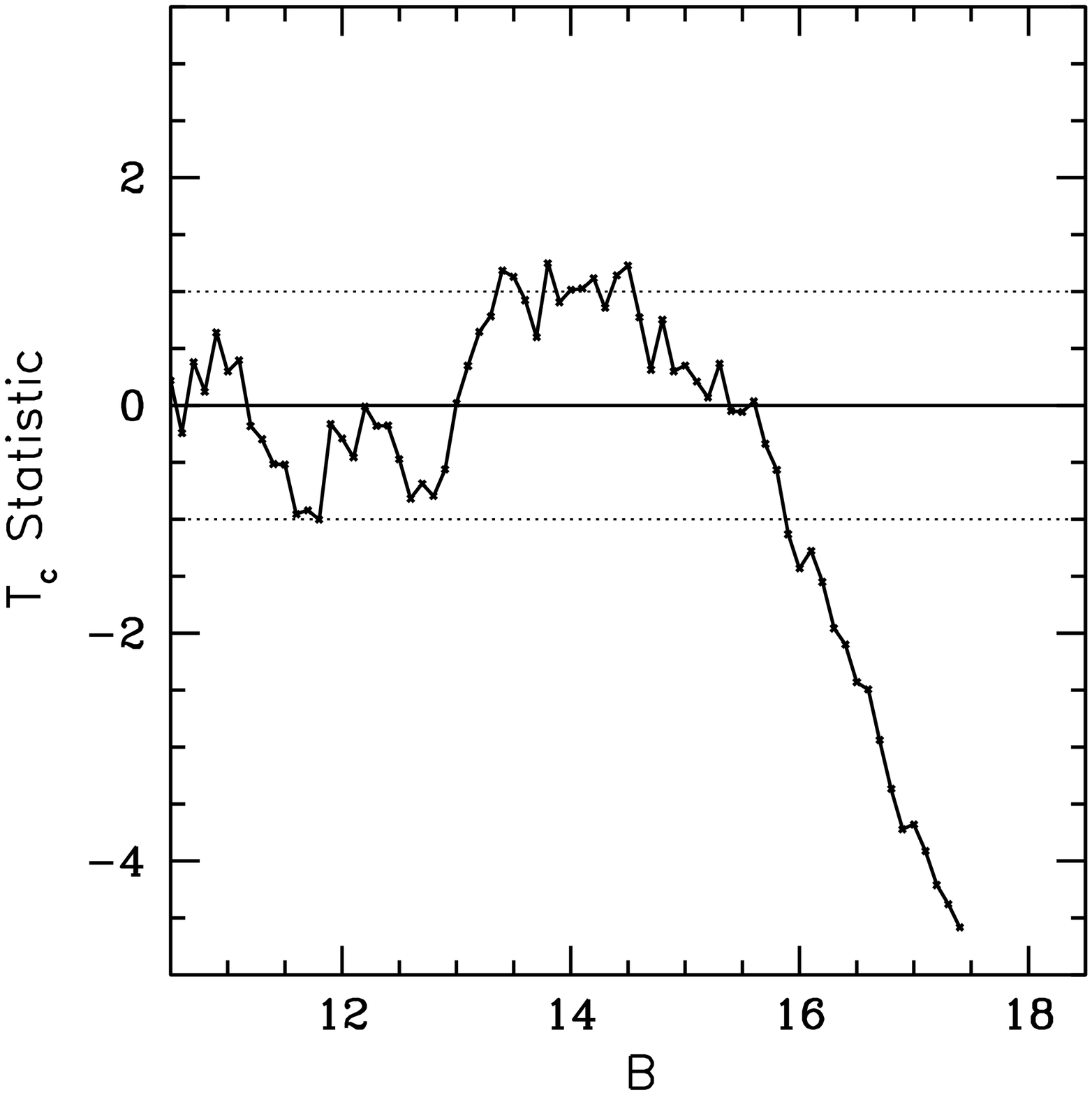}\\
\plotone{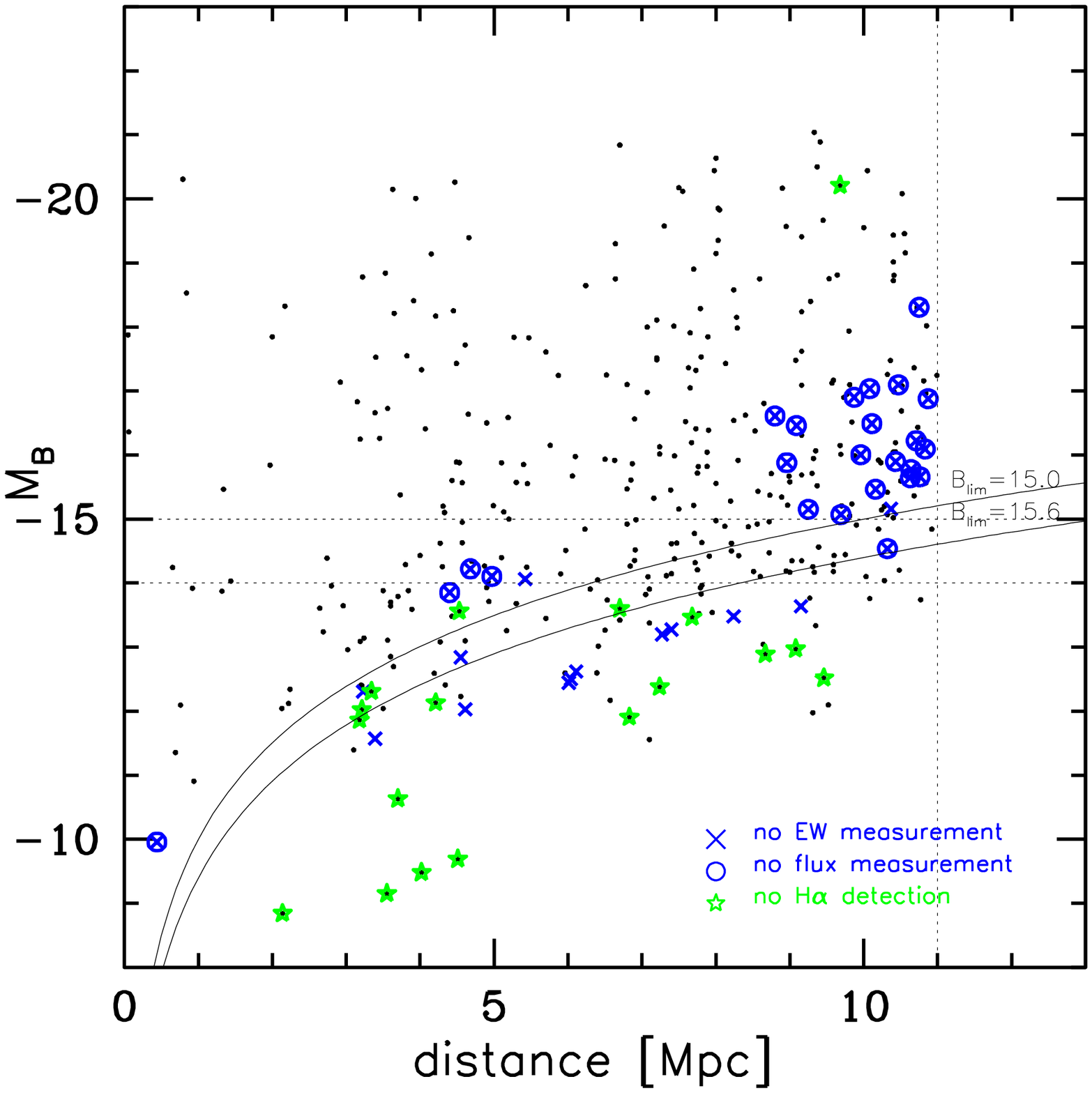}\\
\plotone{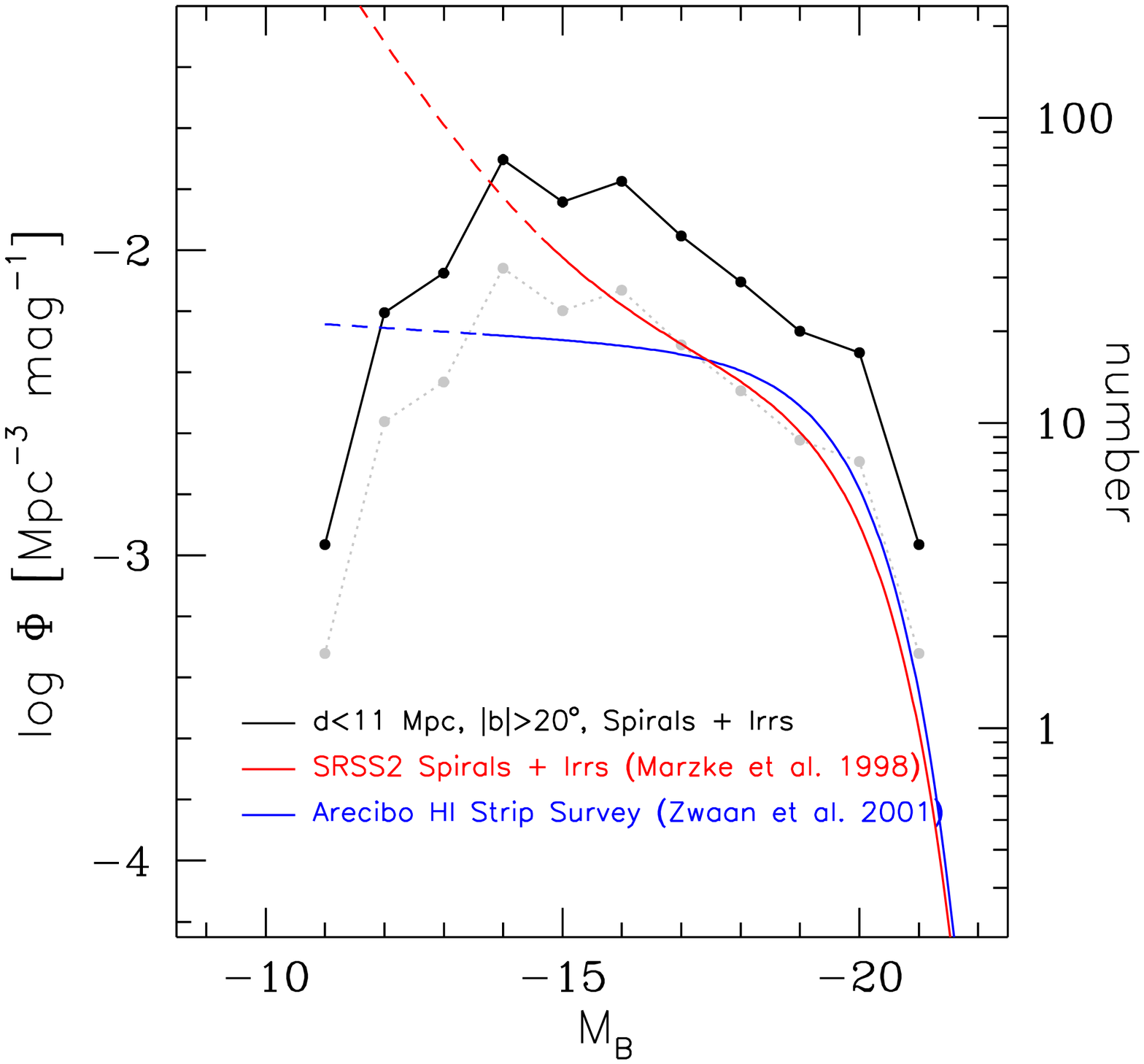}
\vspace*{-0.2in}
\caption{Plots illustrating the completeness of the sample of galaxies within the 11HUGS main survey volume ($|b|>20\degr, d<11$ Mpc). (a) The Rauzy $T_{c}$ statistic as a function of the B-band apparent magnitude corrected for Galactic extinction.  Systematically negative values of $T_c$ indicate that the sample is becoming incomplete.  (b) The B-band absolute magnitude as a function of distance. Non-detections and galaxies without H$\alpha$ measurements are marked as shown in the figure.  (c) Comparison of 11HUGS luminosity densities (black) with luminosity functions (LFs) based on samples with similar morphological make-up from the literature.  The dashed portions of the literature LFs represent extrapolations past the last available data point.  11HUGS luminosity densities renormalized to match the literature luminosity functions are shown in gray.  The right axis indicates the number of galaxies in the 11HUGS sample.  The sample is complete to B=15.6 or $M_B=-14.6$ at 11 Mpc.}  
\end{figure}
\clearpage
\setlength{\voffset}{0mm}

\begin{figure}[p]
\epsscale{0.4}
\plotone{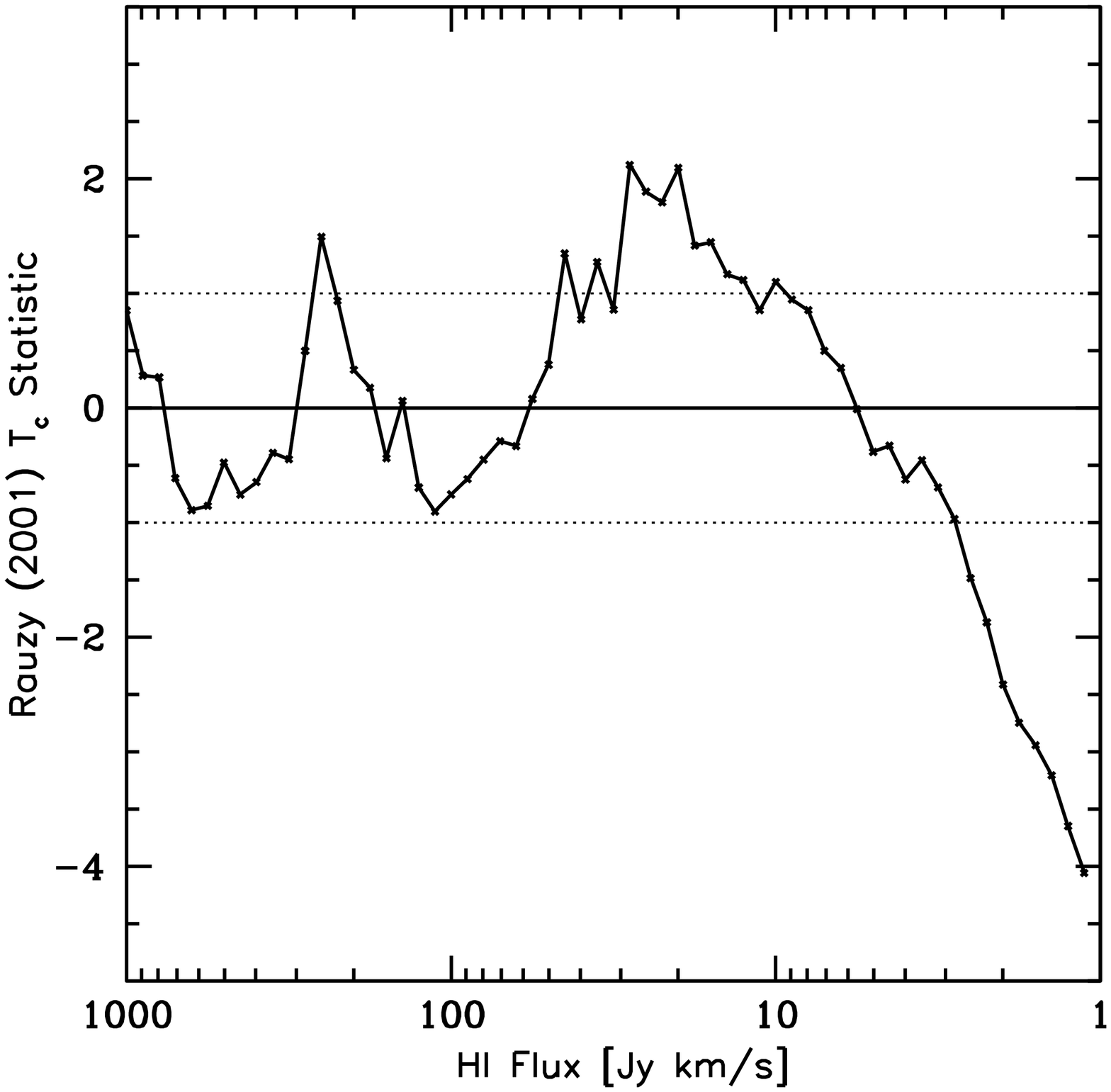}\\
\plotone{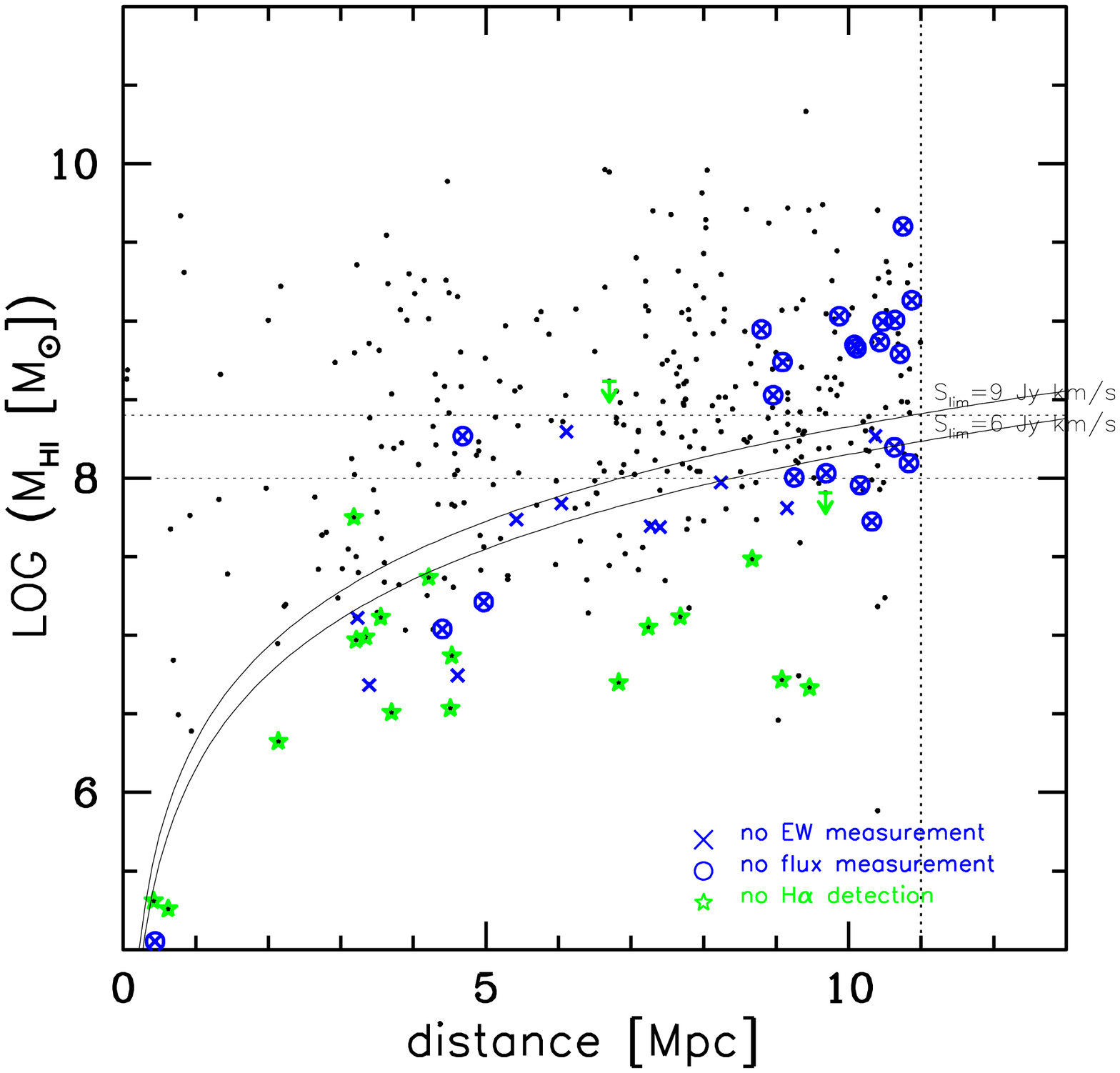}\\
\plotone{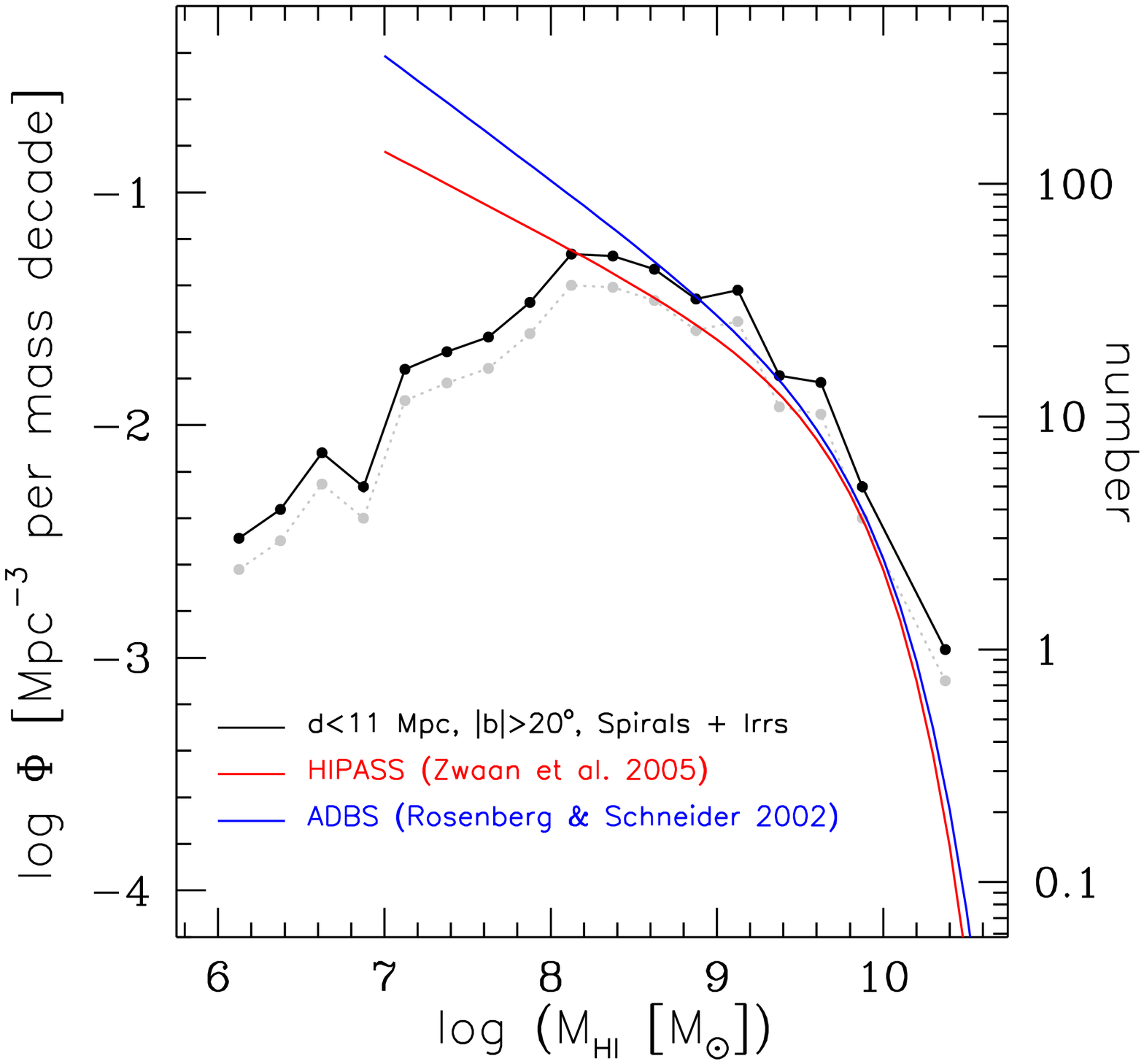}
\caption{Same as Figure 1, but for the 21 cm flux and HI gas mass.  The sample is complete to 6 Jy km s$^{-1}$ or for $M_{HI}>2\times10^8 M_{\odot}$ within our main survey volume ($|b|>20\degr$, $d<11$ Mpc).}
\end{figure}

\begin{figure}[p]
\epsscale{0.4}
\plotone{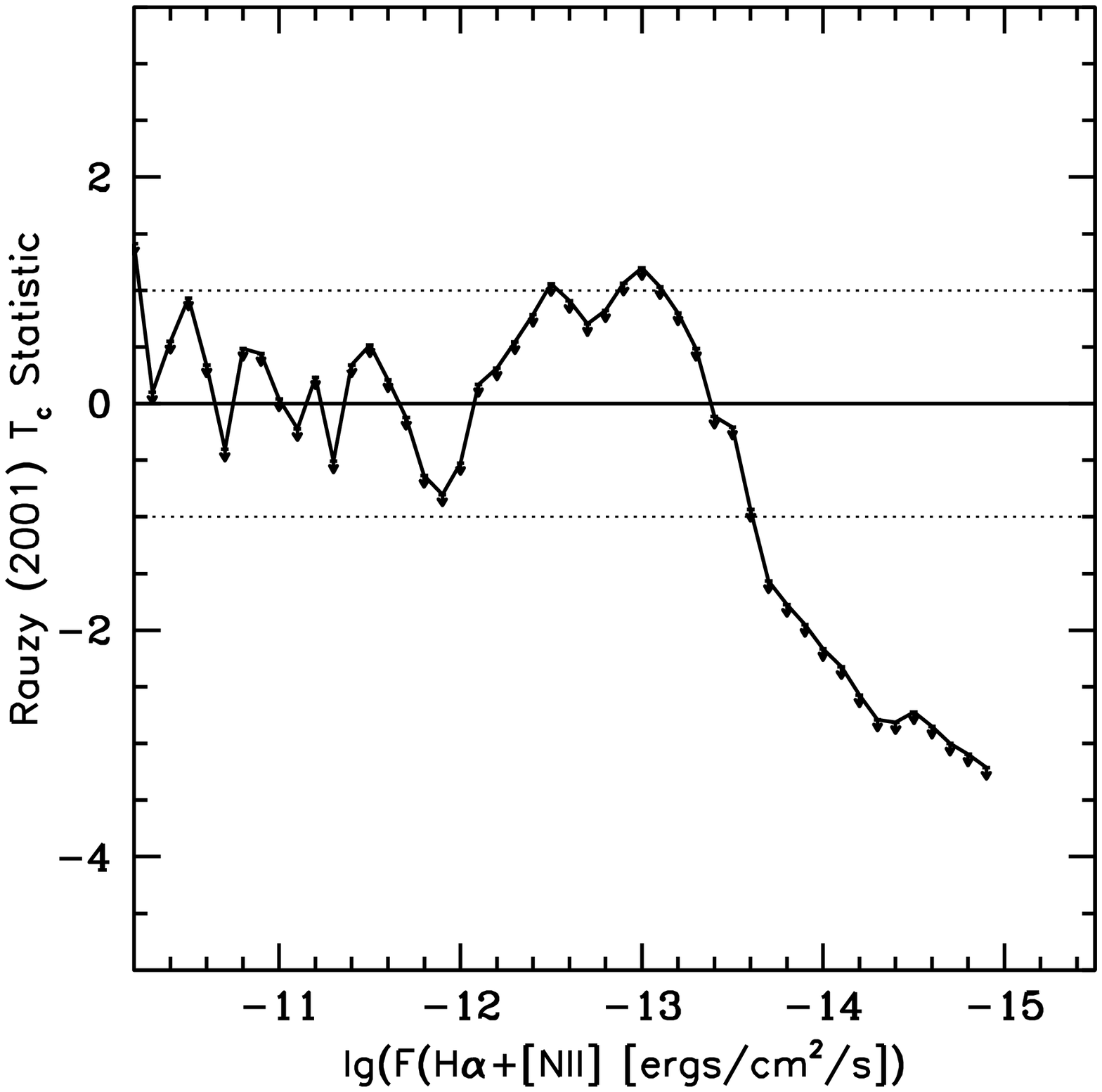}\\
\plotone{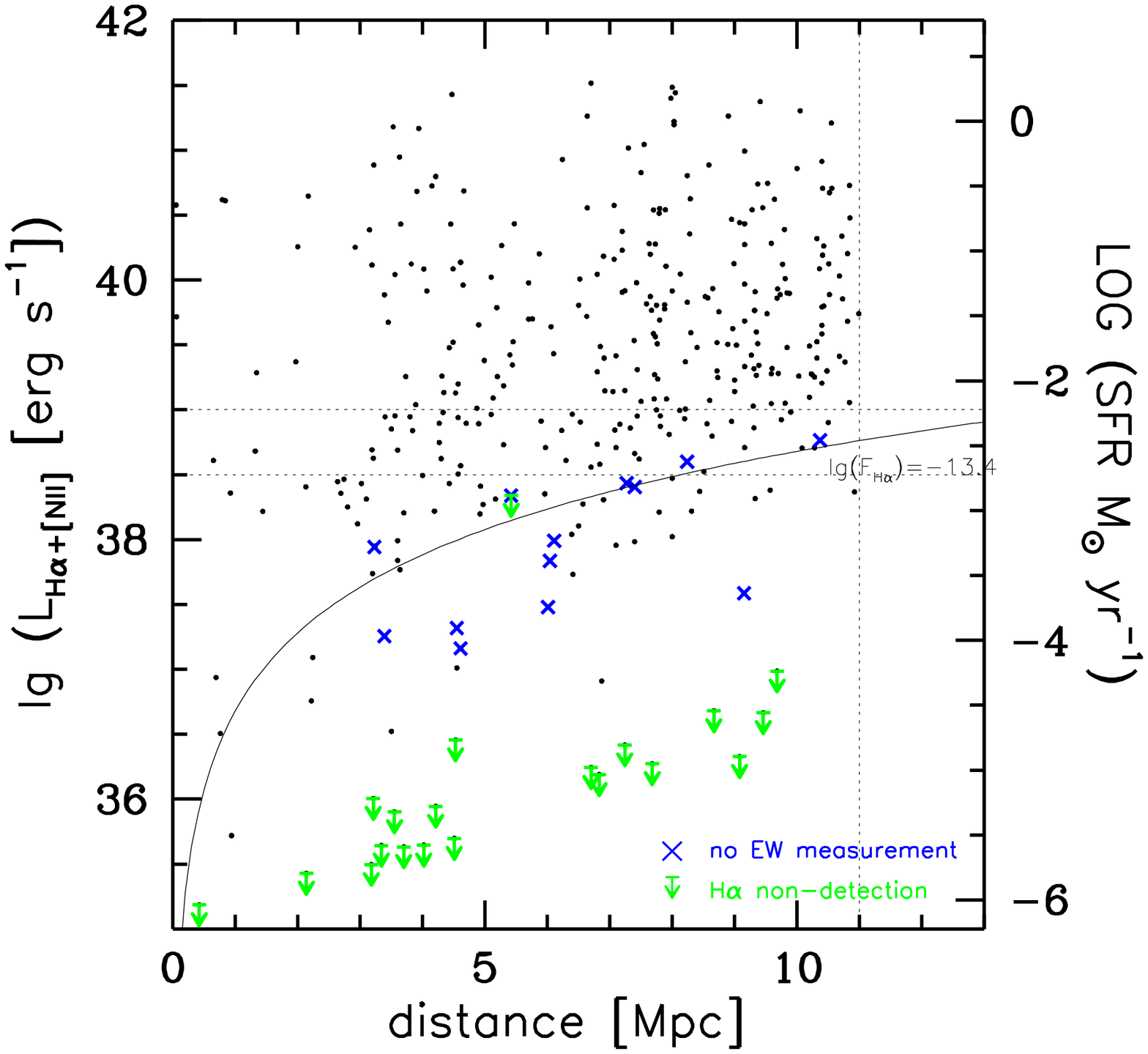}\\
\plotone{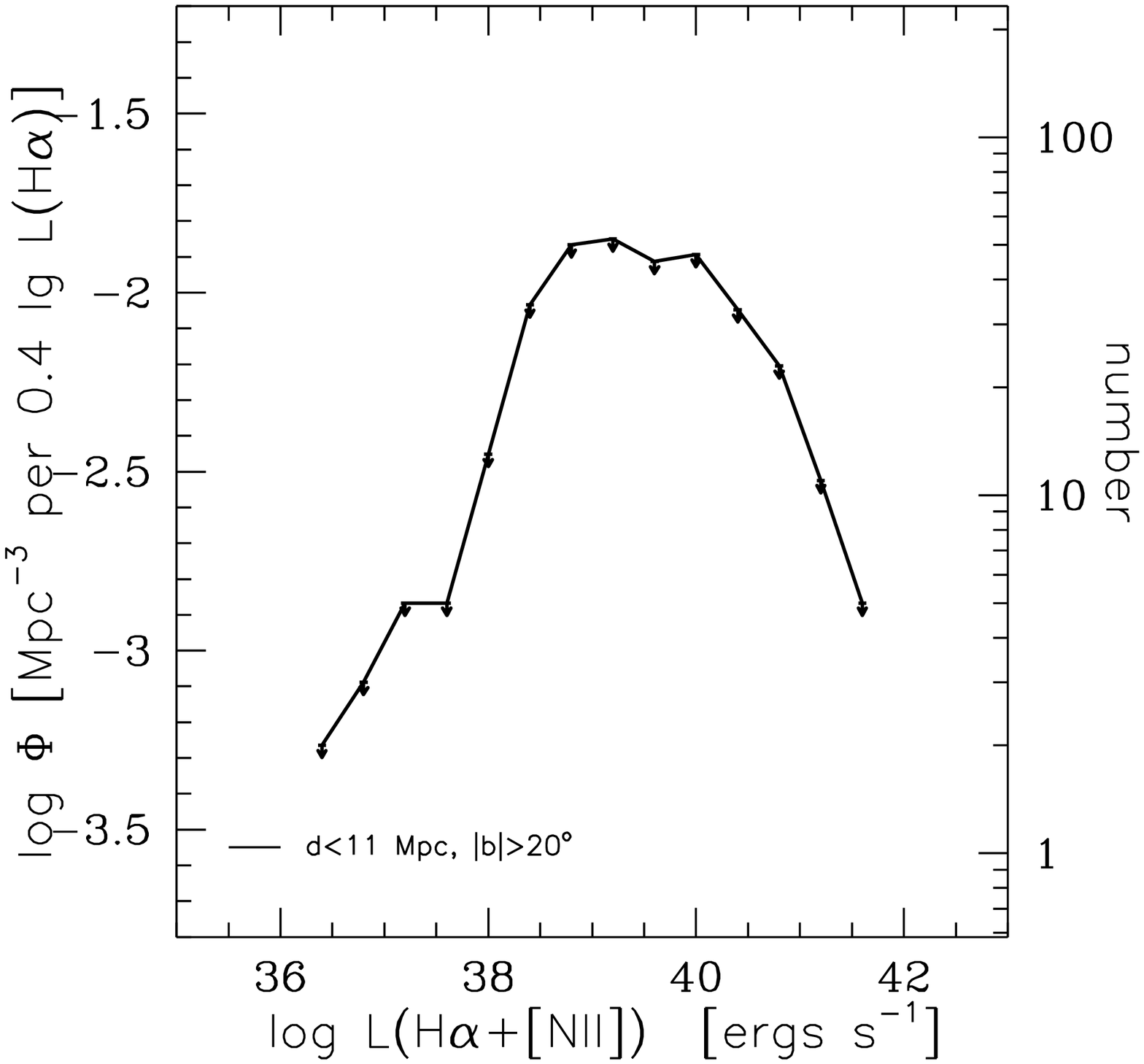}
\caption{Same as Figure 1, but for the H$\alpha$+[NII] flux and luminosity.  The sample is complete to 4 $\times$ 10$^{-14}$ ergs s$^{-1}$ cm$^{-2}$ or  for $L(H\alpha+[NII])>$ 6 $\times$ 10$^{38}$ ergs s$^{-1}$ within our main survey volume ($|b|>20\degr$, $d<11$ Mpc).  No comparable local H$\alpha$ luminosity functions are available to provide secondary checks of completeness at the faint end.}
\end{figure}

\begin{figure}[p]
\epsscale{1}
\plotone{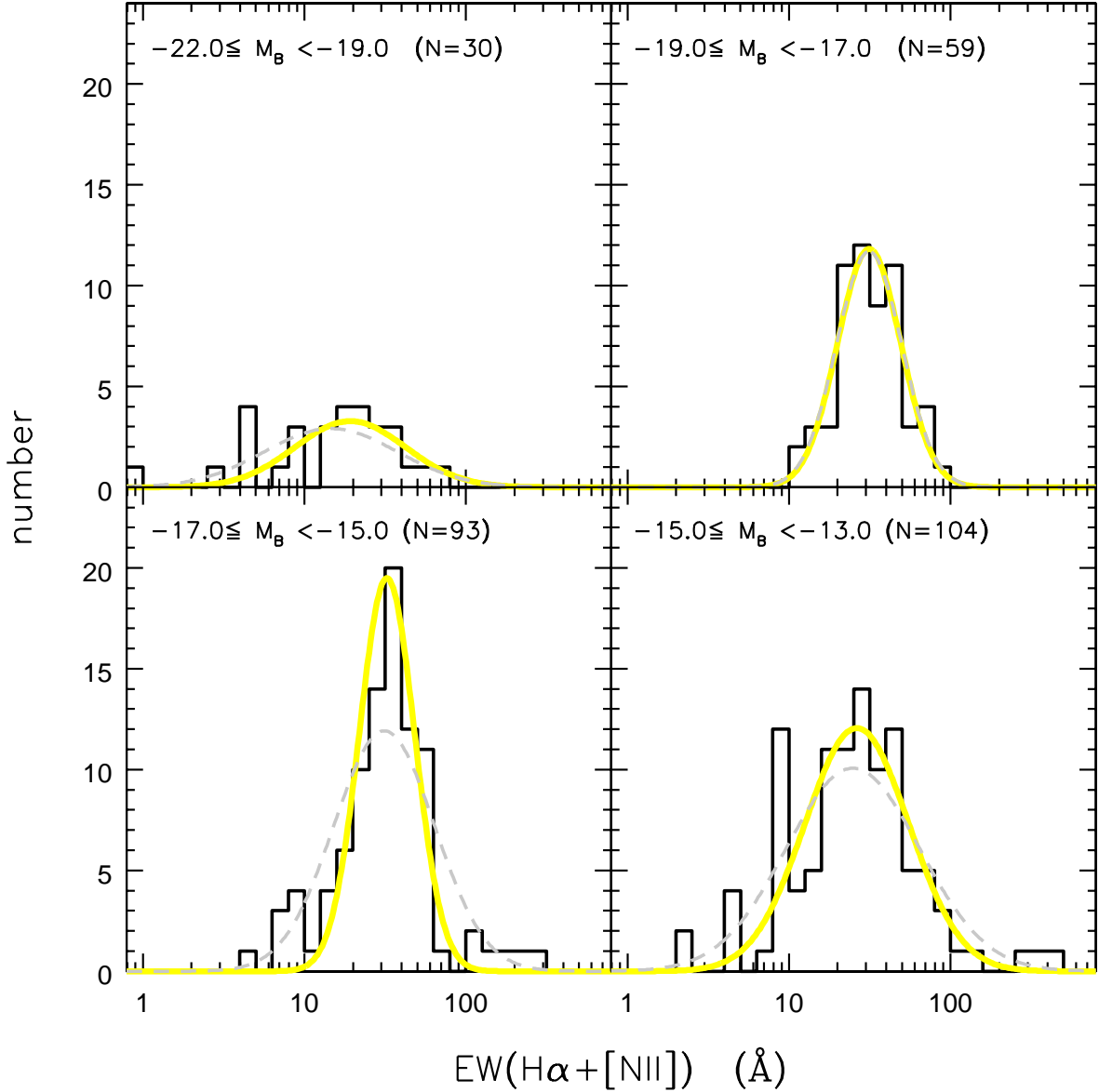}
\caption{Logarithmic H$\alpha$+[NII] EW frequency distributions for galaxies in the 11HUGS main survey volume (black histograms).  The gray dashed curves represent Gaussian functions with means and standard deviations directly computed from the logarithmic EWs in a given luminosity bin, while the best-fit ($\chi^2$ minimized) Gaussian functions to the histograms are shown in yellow.  The integrated EWs of local galaxies can be charaterized by log-normal distributions, but with excess at the tails.   For galaxies with $-19 < M_B < -15$, the correspondence of the upper 3$\sigma$ point of the yellow curves to birthrate parameter values around the starburst threshold ($b\sim$3) suggests that dwarf galaxies currently undergoing global starbursts may be identified as the outliers at the high EW end.  }
\end{figure}

\begin{figure}[p]
\epsscale{1}
\plottwo{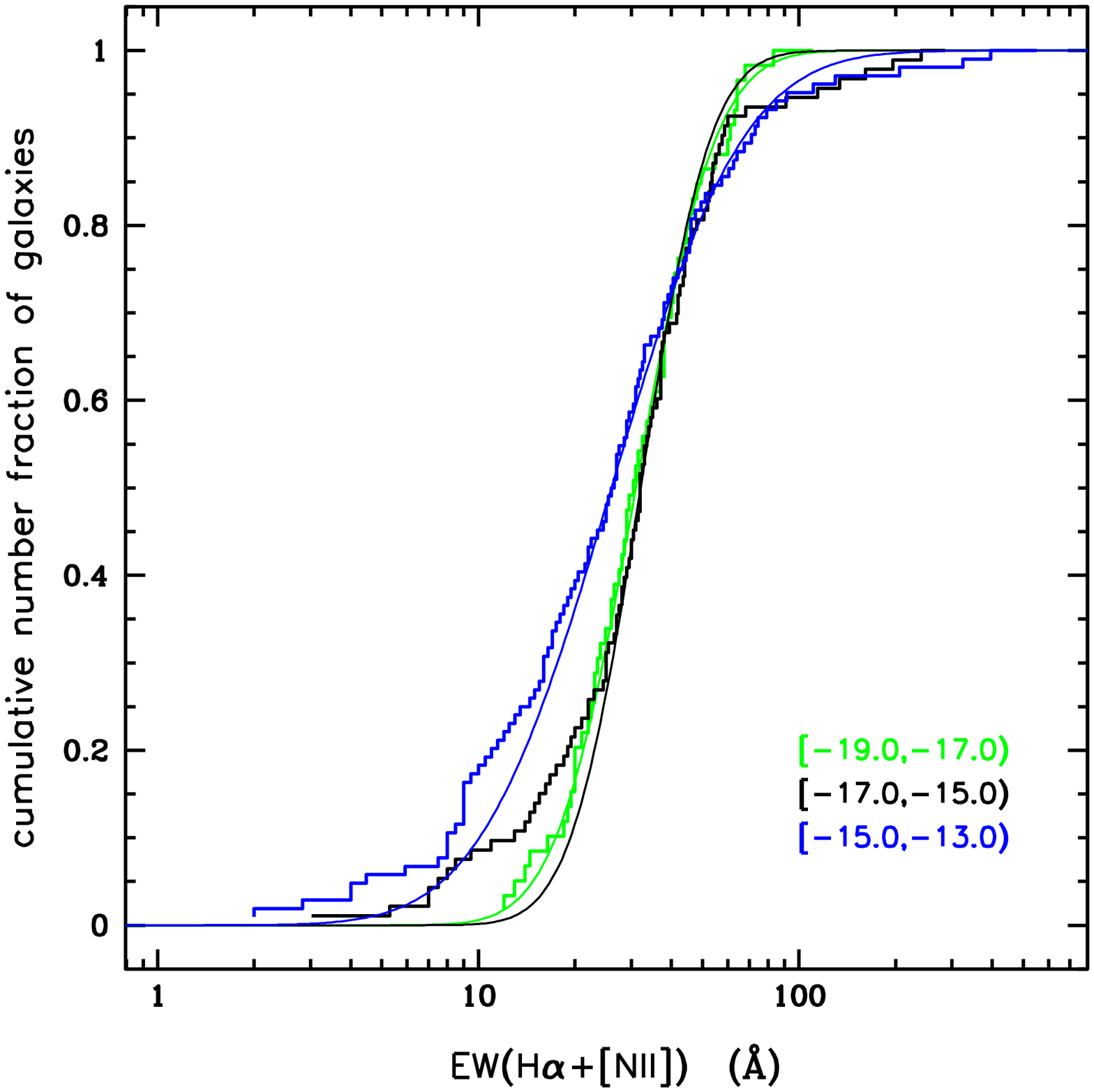}{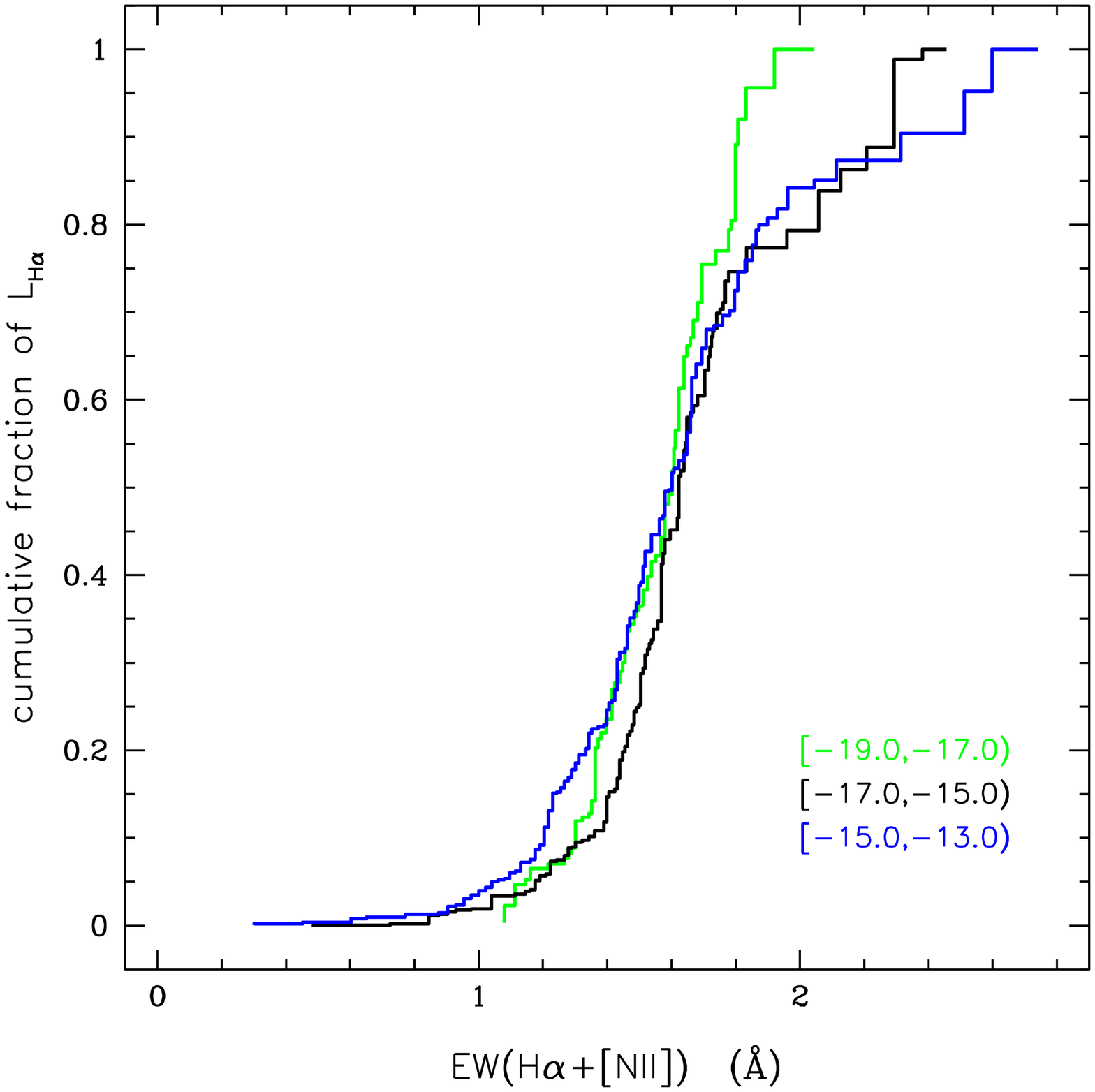}
\caption{(a) Cumulative H$\alpha$+[NII] EW frequency distributions for galaxies in the 11HUGS main survey volume.  The smooth curves show the cumulative distributions of the best-fit Gaussian functions plotted in yellow in Figure 4.  For starbursts defined as objects with EW greater than 100\AA\ ($b\gtrsim2.5$), the number fraction is 6$^{+4}_{-2}$\%.  (b)  The cumulative distribution of L(H$\alpha$+[NII]) as a function of EW(H$\alpha$+[NII]).  About a quarter of the overall star formation in dwarf galaxies occurs in the starburst mode.}
\end{figure}

\begin{figure}[p]
\epsscale{1}
\plotone{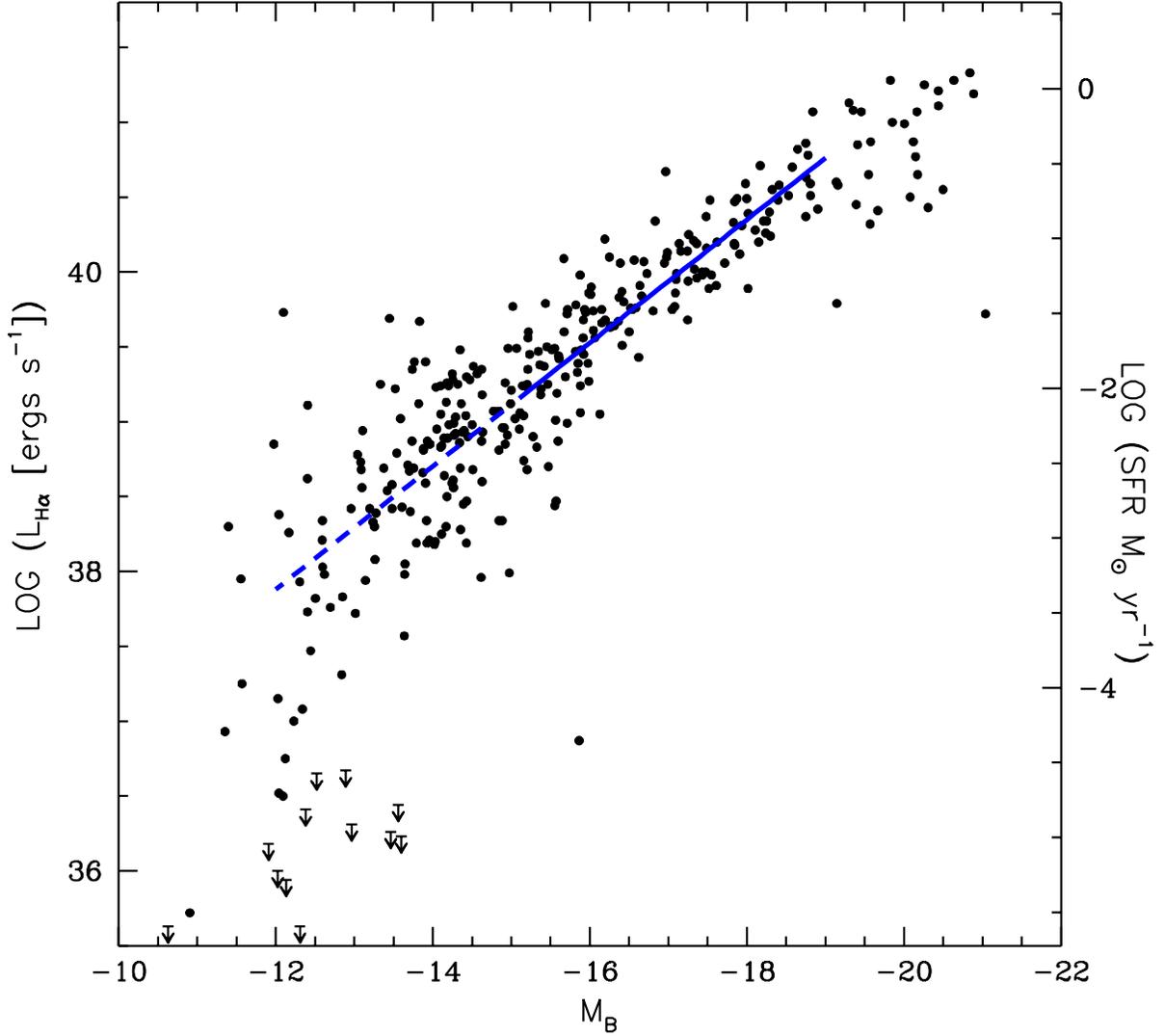}
\caption{$L_{H\alpha}$ vs. $M_B$ for star-forming galaxies with $|b|>20\degr$ and $d<11$ Mpc.  Galactic, but no internal extinction corrections have been applied to the data.  The line represents a fit to the data in the range $-19\leq M_B \leq -15$ which is given by log($L_{H\alpha})=-0.41 M_B+32.94$.  Thus, the H$\alpha$ and $B$-band luminosities roughly follow a linear scaling.  The right axis shows the same SFR scale as used in Figure 3b, and described at the end of \S\ 2.1.1.  Galaxies with $M_B>$-15 should have SFRs high enough to avoid issues with Poisson fluctuations in the formation of high mass ionizing stars.}
\end{figure}

\begin{figure}[p]
\epsscale{0.85}
\plotone{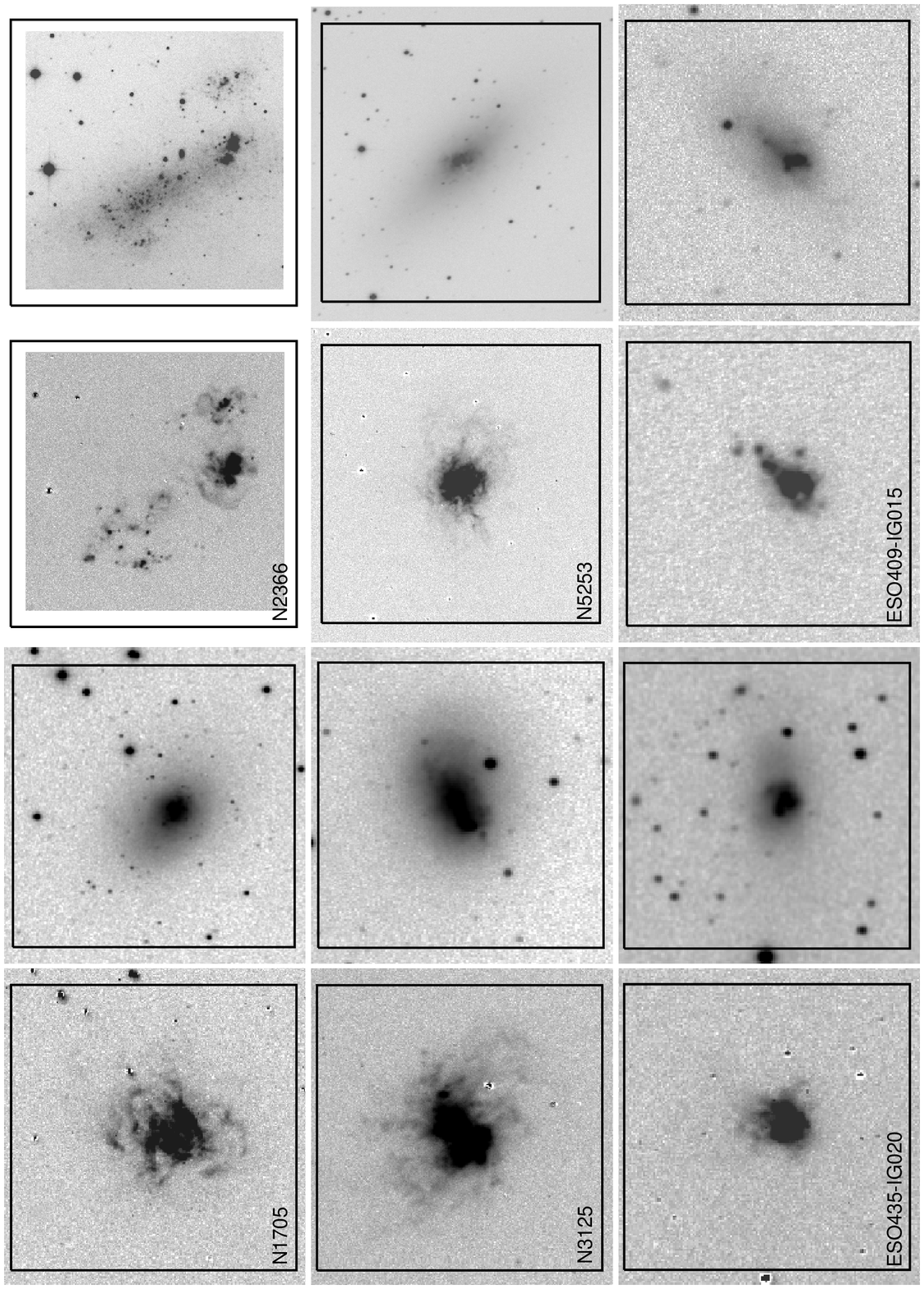}
\caption{Images of dwarf galaxies currently undergoing global starbursts (defined as objects with integrated H$\alpha$ EW greater than 100\AA) in the most robustly populated and complete luminosity bin of the sample  ($-17 \le M_B < -15$).  H$\alpha$+[NII] emission-line only images are shown in the left panels, which are labeled with the galaxy name, while the matching $R$-band images are shown on the right.  The boxes outline 5 $\times$ 5 kpc regions.  The images for ESO409-IG015 are taken from the SINGG survey (Meurer et al. 2006).}
\end{figure}


\begin{thebibliography}{}
\bibitem[]{448} Annibali, F., Greggio, L., Tosi, M., Aloisi, A., \& Leitherer, C. 2003, AJ, 126, 2752
\bibitem[]{449} Aparicio, A., Gallart, G., \& Bertelli, G. 1997a, AJ, 114, 669
\bibitem[]{450} Aparicio, A., Gallart, G., \& Bertelli, G. 1997b, AJ, 114, 680
\bibitem[]{451} Aparicio, A., Tikhonov, N. \& Karachentsev, I. 2000, AJ, 119, 177
\bibitem[]{452} Barazza, F.D., Binggeli, B. \& Prugniel, P. 2001, A\&A, 373, 12
\bibitem[]{453} Bekki, K. 2008, MNRAS, in press
\bibitem[]{454} Bremnes, T., Binggeli, B. \& Prugniel, P. 1998, A\&AS, 129, 313
\bibitem[]{455} Bremnes, T., Binggeli, B. \& Prugniel, P. 2000, A\&AS, 141, 211
n
\bibitem[]{461} Brinchmann, J. et al. 2004, MNRAS, 351, 1151
\bibitem[]{462} Bruzual, G. \& Charlot, S. 2003, MNRAS, 344, 1000
\bibitem[]{467} Cervino, M., Luridiana, V., Perez, E., Vilchez, J. M. \& Valls-Gabaud, D. 2003, A\&A, 407, 177
\bibitem[]{468} Dalcanton, J., et al. 2007, AAS, 211, 7905
\bibitem[]{469} Dekel, A. \& Silk, J., 1986, ApJ, 303, 39
\bibitem[]{471} da Costa et al. 1998, AJ, 116, 1
\bibitem[]{472} de Vaucouleurs, et al. 1991, ``Third Reference Catalogue of Bright Galaxies'' (Berlin, Heidelberg \& New York: Springer-Verlag)
\bibitem[]{473} Dolphin, A.E., Weisz, D.R., Skillman, E.D. \& Holtzman, J.A. 2005, in "Resolved Stellar Populations," eds. D. Valls-Gabaud \& M. Chavez, (San Francisco: ASP)
\bibitem[]{474} Dohm-Palmer, R. C. et al. 1998, AJ, 116, 1227
\bibitem[]{477} Doyle, M.T. et al. 2005, MNRAS, 361, 34
\bibitem[]{484} Gallagher, J.S. \& Hunter, D.A. 1984, ARA\&A, 22, 37
\bibitem[]{485} Gallagher, J.S., Hunter, D.A., \& Tutukov, A.V. 1984, ApJ, 284, 544
\bibitem[]{486} Gallagher, J.S. et al. 1998, AJ, 115, 1869
\bibitem[]{487} Gallagher, J.S. 2005, in ``Starbursts: From 30 Doradus to Lyman Break Galaxies'' ed. R. de Grijs \& R. Gonzalez Delgado (Dordrecht: Springer), 11
\bibitem[]{488} Garnett, D. 2002, ApJ, 581, 1019
\bibitem[]{493} Gil de Paz, A. \& Madore, B. F. 2005, ApJS, 156, 345
\bibitem[]{494} Greggio, L., Marconi, G., Tosi, M. \& Focardi, P. 1993, AJ, 105, 894
\bibitem[]{497} Haro, G. 1956, Bol. Obs. Tonantzintla Tacubaya, 2, 8
\bibitem[]{499} Haynes, M. P., \& Giovanelli, R. 1984, AJ, 89, 758
\bibitem[]{500} Heckman, T.M. 2005, in ``Starbursts: From 30 Doradus to Lyman Break Galaxies,'' ed. R. de Grijs \& R. Gonzalez Delgado (Dordrecht: Springer), 3
\bibitem[]{501} Huchra, J.P. 1977a, ApJS, 35, 171
\bibitem[]{502} Huchra, J.P. 1977b, ApJ, 217, 928
\bibitem[]{504} Hunter, D.A. \& Gallagher, J.S., 1985, ApJS, 58, 533
\bibitem[]{505} Hunter, D.A. \& Gallagher, J.S., 1986, PASP, 98, 599
\bibitem[]{506} Hunter, D.A. \& Elmegreen, B.G, 2004, AJ, 128, 2170
\bibitem[]{508} James, P.A. et al. 2004, A\&A, 414, 23
\bibitem[]{509} James, P. A., Knapen, J. H., Shane, N. S., Baldry, I. K. \& de Jong, R. S. 2008, A\&A, 482, 507
\bibitem[]{514} Kennicutt, R.C., Tamblyn, P. \& Congdon, C.E. 1994, ApJ, 435, 22
\bibitem[]{515} Kauffmann et al., 2003a, MNRAS, 341, 33
\bibitem[]{516} Kauffmann et al., 2003b, MNRAS, 341, 54
\bibitem[]{517} Kennicutt, R.C. 1998, ARA\&A, 36, 189
\bibitem[]{518} Kennicutt, R.C., Lee, J.C., Funes, J.G., Sakai, S. \& Akiyama, S. 2008, ApJS, in press (Paper I)

\bibitem[]{521} Kennicutt, R.C., Lee, J.C., Funes, J.G., Sakai, S. \& Akiyama, S. 2005, in "Starbursts: From 30 Doradus to Lyman Break Galaxies," ed. R. de Grijs \& R. Gonzalez Delgado (Dordrecht: Springer), 187

\bibitem[]{525} Karachentsev, I.D., Karachentseva, V.E., Huchtmeier, W.K. \& Makarov, D.I. 2004, AJ, 127, 2031
\bibitem[]{530} Lasker, B.M. et al. 1990, AJ, 99, 2019
\bibitem[]{531} Lee, H. 2006, ApJ, 647, 970
\bibitem[]{532} Lee, J.C., 2006, Ph.D. thesis, Univ. Arizona
\bibitem[]{533} Lee, J.C., Salzer, J.J., Impey, C., Thuan, T. X. \& Gronwall, C. 2002, AJ, 124, 3088
\bibitem[]{534} Lee, J.C., Salzer, J.J. \& Melbourne, J.  2004, ApJ, 616, 752
\bibitem[]{535} Lee, J.C. et al. 2008, in "Formation and Evolution of Galaxy Disks," ed. J. Funes 
\bibitem[]{536} Lee, J.C., Kennicutt, R.C., Funes, J., Sakai, S. \& Akiyama, S. 2007, ApJ, 671L, 113 
\bibitem[]{540} Longair, M.S. et al. 1998, "Galaxy Formation" (Berlin: Springer-Verlag)
\bibitem[]{541} Mac Low, M.-M. \& Ferrara, A. 1999, ApJ, 513, 142
\bibitem[]{543} Marconi, G., Tosi, M., Greggio, L. \& Focardi, P. 1995, AJ, 109, 173
\bibitem[]{544} Marlowe, A.T., Heckman, T.M., Wyse, R.F.G. \&  Schommer, R. 1995, ApJ, 438, 563
\bibitem[]{545} Marlowe, A.T., Meurer, G.R. \& Heckman, T.M. 1999, ApJ, 522, 183
\bibitem[]{546} Markarian, B.E. 1967, Astrofizika, 3, 55
\bibitem[]{547} Markarian, B.E. 1969a, Astrofizika, 5, 443
\bibitem[]{548} Markarian, B.E. 1969b, Astrofizika, 5, 581
\bibitem[]{549} Martin, C.L. 1999, ApJ, 513, 156
\bibitem[]{550} Marzke, R.O. et al. 1998, ApJ, 503, 617
\bibitem[]{553} Meurer, G. et al. 2006, ApJS, 165, 307
\bibitem[]{554} Meyer, M.J. et al. 2004, MNRAS, 350, 1195
\bibitem[]{555} Miller, B.W. \& Hodge, P. 1994, ApJ, 427, 656
\bibitem[]{560} Nilson, P. 1973, Uppsala General Catalogue of Galaxies (Uppsala: Uppsala Astron. Obs.)
\bibitem[]{565} Oey, M.S. \& Kennicutt, R.C. 1997, MNRAS, 291, 827
\bibitem[]{566} Papaderos, P., Loose, H.-H., Fricke, K. J., \& Thuan, T. X. 1996, A\&A, 120, 207
\bibitem[]{567} Parodi, B. R., Barazza, F. D. \& Binggeli, B. 2002, A\&A, 388, 29
\bibitem[]{568} Paturel, G. et al. 2003, A\&A, 412, 57
\bibitem[]{569} Pelupessy, F. I., van der Werf, P. P. \& Icke, V. 2004, A\&A, 422, 55
\bibitem[]{573} Rauzy, S. 2001, MNRAS, 324, 51
\bibitem[]{575} Richer, M. G., \& McCall, M. L. 1995, ApJ, 445, 642
\bibitem[]{578} Rosenberg, J.L. \& Schneider, S.E. 2002, ApJ, 567, 247
\bibitem[]{580} Salzer, J.J. 1989, ApJ, 347, 152
\bibitem[]{581} Salzer, J.J. et al. 2000, AJ, 120, 80
\bibitem[]{582} Salzer, J.J. et al. 2001, AJ, 121, 66
\bibitem[]{583} Sargent, W.L.W. 1972, ApJ, 173, 7
\bibitem[]{584} Sargent, W.L.W. \& Searle, L. 1970, ApJ, 162, 155
\bibitem[]{587} Schmidt, M. 1968, ApJ, 151, 393
\bibitem[]{588} Searle, L. \& Sargent, W.L.W., 1972, ApJ, 173, 25
\bibitem[]{589} Searle, L., Sargent, W.L.W. \& Bagnuolo, W.G. 1973, ApJ, 179, 427
\bibitem[]{591} Simpson, C.E. \& Gottesman, S. T. 2000, AJ, 120, 2975
\bibitem[]{592} Skillman, E. D., Kennicutt, R. C. \& Hodge, P. W. 1989, ApJ, 347, 875
\bibitem[]{593} Skillman, E.D. \&  Bender, R., 1995, RMxAC, 3, 25
\bibitem[]{594} Stinson, G. S., et al. 2007, ApJ, 667, 170
\bibitem[]{596} Springob, C.M., Haynes, M.P., Giovanelli, R. \& Kent, B. 2005, ApJS, 160, 149
\bibitem[]{599} Tinsley, B.M. 1968, ApJ, 151, 547
\bibitem[]{600} Tinsley, B.M. 1972, A\&A, 20, 383
\bibitem[]{602} Tosi, M., Greggio, L., Marconi, G. \& Focardi, P. 1991, AJ, 102, 951
\bibitem[]{603} Tremonti, C.A. et al. 2004, ApJ, 613, 898
\bibitem[]{604} Tremonti, C.A., Lee, J.C., van Zee, L., Kennicutt, R.C., Gil de Paz, A., Sakai, S. \& Funes, J. in prep.
\bibitem[]{607} Tully, R.B. 1988a, ``Nearby Galaxies Catalog'' (Cambridge \& New York: Cambridge University Press)
\bibitem[]{608} Tully, R.B. 1988b, AJ, 96, 73
\bibitem[]{611} Vader, J.P. 1986, ApJ, 305, 669
\bibitem[]{613} van Zee, L., Haynes, M. P., Salzer, J. J., \& Broeils, A. H. 1996, AJ, 112, 129
\bibitem[]{614} van Zee, L., Haynes, M.P. \& Salzer, J.J. 1997, AJ, 114, 2479
\bibitem[]{616} van Zee, L. 2000, AJ, 119, 2757
\bibitem[]{617} van Zee, L. 2001, AJ, 121, 2003
\bibitem[]{618} Weisz, D., et al. 2008, ApJ, submitted
\bibitem[]{622} Weidner, C. \& Kroupa, P. 2006, MNRAS, 365, 1333
\bibitem[]{624} Zamorano, J. et al. 1994, ApJS, 95, 387

\bibitem[]{628} Zwaan, M.A., Briggs, F.H., Sprayberry, D. \& Sorar, E. 1997, ApJ, 490, 173
\bibitem[]{629} Zwaan, M.A., Briggs, F.H. \& Sprayberry, D. 2001, MNRAS, 327, 1249
\bibitem[]{630} Zwaan, M.A. et al. 2003, AJ, 125, 2842
\bibitem[]{631} Zwaan, M. A., Meyer, M. J., Staveley-Smith, L. \& Webster, R. L. 2005, MNRAS, 359, 30
\bibitem[]{632} Zwicky, F. 1964, ApJ, 140, 1467
\bibitem[]{633} Zwicky, F. 1966, ApJ, 143, 192
\bibitem[]{634} Zwicky, F. \& Zwicky, M.A. 1971, Catalogue of selected compact galaxies and of post-eruptive galaxies (Guemligen: Zwicky)
\bibitem[]{635} Zwicky, F., Herzog, E., Kowal, C.T., Wild, P. \& Karpowicz, M. 1961-1968, Catalogue of Galaxies and of Clusters of Galaxies (Pasadena: Caltech)

\end{thebibliography}
\end{document}